\documentclass[twocolumn,twocolappendix]{aastex631}
\usepackage{amsmath} 
\usepackage{graphicx}

\newcommand*{\hi}{\rm{H}\,\rm{\textsc{i}}}
\newcommand*{\msun}{\ensuremath{\rm{M}_{\odot}}}

\newcommand*{\kms}{\text{km}\,\text{s}\ensuremath{^{-1}}}

\begin{document}

\title{The Disturbed and Globular Cluster-Rich Ultra-diffuse Galaxy UGC 9050-Dw1}

\correspondingauthor{Catherine E. Fielder}
\email{cfielder@arizona.edu}

\author[0000-0001-8245-779X]{Catherine E. Fielder}
\affiliation{Steward Observatory, University of Arizona, 933 North Cherry Avenue, Tucson, AZ 85721-0065, USA}

\author[0000-0002-5434-4904]{Michael G. Jones}
\affiliation{Steward Observatory, University of Arizona, 933 North Cherry Avenue, Rm. N204, Tucson, AZ 85721-0065, USA}

\author[0000-0003-4102-380X]{David J. Sand}
\affiliation{Steward Observatory, University of Arizona, 933 North Cherry Avenue, Rm. N204, Tucson, AZ 85721-0065, USA}

\author[0000-0001-8354-7279]{Paul Bennet}
\affiliation{Space Telescope Science Institute, 3700 San Martin Drive, Baltimore, MD 21218, USA}

\author[0000-0002-1763-4128]{Denija Crnojevi\'{c}}
\affil{Department of Physics and Astronomy, University of Tampa, 401 West Kennedy Boulevard, Tampa, FL 33606, USA}

\author[0000-0001-8855-3635]{Ananthan Karunakaran}
\affiliation{Instituto de Astrof\'{i}sica de Andaluc\'{i}a (CSIC), Glorieta de la Astronom\'{i}a, 18008 Granada, Spain}

\author[0000-0001-9649-4815]{Bur\c{c}in Mutlu-Pakdil}
\affil{Department of Physics and Astronomy, Dartmouth College, Hanover, NH 03755, USA}

\author[0000-0002-0956-7949]{Kristine Spekkens}
\affiliation{Department of Physics and Space Science, Royal Military College of Canada P.O. Box 17000, Station Forces Kingston, ON K7K 7B4, Canada}
\affiliation{Department of Physics, Engineering Physics and Astronomy, Queen’s University, Kingston, ON K7L 3N6, Canada}



\begin{abstract}

We investigate the ultra-diffuse galaxy (UDG) UGC~9050-Dw1, which was selected because of its disturbed morphology as part of a larger sample of UDGs that display evidence for significant interactions. We use the Hubble Space Telescope's Advanced Camera for Surveys to identify globular clusters (GCs) associated with UGC~9050-Dw1, and the Jansky Very Large Array to measure its \hi\ content.  UGC~9050-Dw1, a neighbor to the low surface brightness spiral UGC~9050, exhibits a unique UV bright central ``clump'' with clearly associated \hi\ gas and an extended stellar tidal plume to the north. We identify $52\pm12$ GCs, implying a specific frequency $S_\mathrm{N} = 122\pm38$, one of the highest reported for a UDG of this luminosity ($\log{\rm{L}_{V}/\rm{L}_{\odot}} = 7.5\pm0.1$). Additionally, $\sim 20\%$ of the total light of the galaxy is contributed by GCs. Nearly uniform GC colors suggest they were formed during a single intense episode of star formation. We posit that UGC~9050-Dw1 represents the initial definitive observational example of UDG formation resulting from a dwarf merger event, where subsequent clumpy star formation has contributed to its present observed characteristics. 

\end{abstract}



\section{Introduction} 
\label{sec:intro}

Ultra diffuse galaxies (UDGs) have garnered a lot of interest since the discovery of large populations of extreme low surface brightness (LSB) galaxies in the Coma \citep[e.g.,][]{koda2015,vandokkum2015,yagi2016,amorisco2018,lim2018,forbes2020}, Virgo \citep[e.g.,][]{mihos2015,lim2020}, Perseus \citep[e.g.,][]{wittmann2017}, and Fornax \citep[e.g.,][]{venhola2017} clusters. A flurry of additional studies and searches ensued, in an attempt to understand these large (half-light radii $>1.5$ kpc), LSB objects (central $g$-band surface brightness $>24$ mag arcsec$^{-2}$). UDGs are remarkable, with stellar masses similar to those expected for dwarf galaxies, yet with physical sizes comparable to the Milky Way (\citealt{vandokkum2015}, see discussion on size in \citealt{chamba2020}). In addition, UDGs have been found in abundance across all environments in addition to clusters, including 
groups \citep{merritt2016,bennet2017,roman2017,vanderburg2017,spekkens2018,jones2021}, and in the field \citep{martinez2016,leisman2017,greco2018,janowiecki2019,roman2019,prole2019,prole2021,jones2023}. The ubiquity of UDGs across all environments, and their extreme nature, have challenged galaxy formation and evolution models.

A number of potential formation mechanisms for UDGs have been proposed, and it is likely that multiple formation pathways are necessary \citep[e.g.,][]{pandya2018}. Some explanations for the extended, low surface brightness nature of UDGs include strong stellar feedback \citep{dicintio2017,chan2018}, high-spin dark matter (DM) halos \citep{amorisco2016,rong2017}, early mergers \citep{wright2021}, or failed galaxies that reside in exceptionally massive DM halos \citep{vandokkum2016}. Others are likely `puffed up' dwarfs \citep{conselice2018,bennet2018,carleton2019,tremmel2020,jones2021}, or the result of gas rich galaxy collisions yielding tidal dwarf galaxies (TDGs; e.g., \citealt{duc1998,duc2012,roman2021}). Some authors also suggest that the observed UDG population may be the result of a combination of several of these formation scenarios \citep[e.g.,][]{ruiz-lara2019,trujillo2019}.


One of the most important clues for understanding the origins of UDGs comes from their globular cluster (GC) population. Old GCs trace the early epochs of galaxy assembly \citep[see e.g.,][and references therein]{kissler-patig2000}, making their abundance an excellent discriminator in constraining UDG origins. Some studies have found UDGs with much more abundant GC populations than would be expected for a dwarf-mass dark matter halo, while others find the GC systems of many UDGs are consistent with dwarf-mass halos \citep{beasley2016A,roman2017,amorisco2018,lim2020,somalwar2020,saifollahi2021}. Additionally, GC abundance has been found to be strongly correlated with the total system mass - particularly DM halo mass \citep[e.g.,][]{harris2013,prole2019,zaritsky2022} - which is also true for the UDG regime \citep{harris2017} allowing for DM halo mass constraints of UDGs to be photometrically obtained. Spectroscopic studies of GCs in UDGs can determine velocity gradients to constrain the evolutionary histories of UDGs further via signals of tidal disruption \citep{beasley2016,toloba2018,toloba2023}. Consequently GC abundance has often been used as a means for discriminating between potential formation scenarios when comparing sub-populations of UDGs that otherwise are quite similar in luminosity and stellar mass. 

For UDGs with evidence of ongoing star formation, neutral gas (\hi) observations provide additional, complementary information. \hi\ is typically one of the most loosely bound baryonic components of a galaxy and therefore is a remarkably sensitive tracer of tidal interactions. Thus, the \hi\  morphology of a UDG, or even the absence of neutral gas \citep{jones2021}, can be used to narrow down potential formation scenarios. \hi \ line widths and velocity maps also provide insight into the bulk kinematics of a galaxy, and have been used to argue that some \hi-rich UDGs might be DM poor \citep{mancera2019,mancera2022}.


In this work we focus on a UDG identified by a semi-automated search \citep{bennet2017} of the Canada–France–Hawaii Telescope Legacy Survey (CFHTLS): UGC~9050-Dw1. This UDG plausibly resides in a group environment with the low surface brightness spiral UGC~9050 (see e.g., \citealt{pahwa2018}), and is detected in the UV. Only a handful of such group UDGs with indication of recent star formation have been studied \citep[e.g.,][]{roman2017}. In the CFHT data, the UDG has an evident tail-like feature and the central region exhibits an unusual morphology, indicative of an interaction. We use both Hubble Space Telescope (HST) and Jansky Very Large Array (VLA) observations to identify GC candidates and measure the \hi\ morphology of UGC~9050-Dw1 in order to distinguish between the variety of formation scenarios befitting a UDG with evidence for tidal disturbances. 

This paper is outlined in the following manner. \autoref{sec:data} provides an overview of the VLA and HST data, in addition to other ancillary data used in this work. In \autoref{sec:properties} we detail the derived properties of UGC~9050-Dw1. \autoref{sec:gcc_selection} describes the criteria for selecting globular cluster candidates and \autoref{sec:gc_props} details the GC abundance, properties, and inferred halo mass. In \autoref{sec:discussion} we discuss possible formation mechanisms for UGC~9050-Dw1. Finally, \autoref{sec:conclusion} provides a summary of our findings and conclusions. \autoref{sec:appendix} provides supplementary tables and figures, including a full table of GC candidates. All photometry used in this work is in the Vega magnitude system unless otherwise stated. 

\section{Observational Data} 
\label{sec:data}


UGC~9050-Dw1 was identified in ground-based CFHTLS imaging by a semi-automated search for diffuse dwarfs (initial results are presented in \citealt{bennet2017}). This now completed search covers $\sim150\;\rm{deg}^{2}$, within which hundreds of diffuse dwarf and UDG candidates have been identified and several UDGs have been confirmed. \citet{bennet2018} initially reported two UDGs as part of this automated search, which were subsequently followed up with HST and VLA observations in \citet{jones2021}. These two UDGs were selected due to their apparent association with tidal streams, part of a larger case study for constraining formation mechanisms for UDGs with plausible evidence for interactions with larger galaxies. These UDGs specifically had little evidence of star formation or neutral gas. UGC~9050-Dw1 was  selected as an extension to this study, focusing on UDGs with signs of interaction and  evidence of recent star formation. For the case of UGC~9050-Dw1 this  is apparent because of its blue color and associated GALEX \citep[GAlaxy Evolution EXplorer;][]{galex2005} NUV emission. 

UGC~9050 is UGC~9050-Dw1's nearest apparent neighbor and presumed host galaxy. The two are separated by less than 50~km/s in radial velocity (see \autoref{subsec:hi_props}). We adopt the distance of UGC~9050 as measured from the Virgo Infall Hubble flow \citep{mould2000} ($H_{0}=67.8$~km/sec/Mpc) of $35.2\pm2.5$ Mpc, which we also adopt for UGC~9050-Dw1 given their extremely similar recessional velocity. At the adopted distance the two are separated by $69$~kpc in projection. 
UGC~9050-Dw1 and UGC~9050 also both lie close in proximity ($\sim 300$ kpc) and within redshift space ($\Delta v_{\rm{helio}}<100$ km/s) of the NGC 5480 and 5481 pair, indicating that they may also be fringe members of the NGC~5481 group ($D = 35\pm2.5$~kpc). 

\subsection{HST Observations}
\label{subsec:hst_obs}

UGC~9050-Dw1 was observed in September of 2022 under HST program ID 16890 \citep{sand2021}. This target was observed with the Advanced Camera for Surveys (ACS) with the Wide Field Channel (WFC). Observations were completed in the F555W and F814W filters, with 2406~s and 2439~s exposure times, respectively. Additionally, Wide Field Camera 3 (WFC3) images were taken in parallel to use as a nearby reference background field. 

\begin{figure*}
    \centering
    \includegraphics[width=0.75\linewidth]{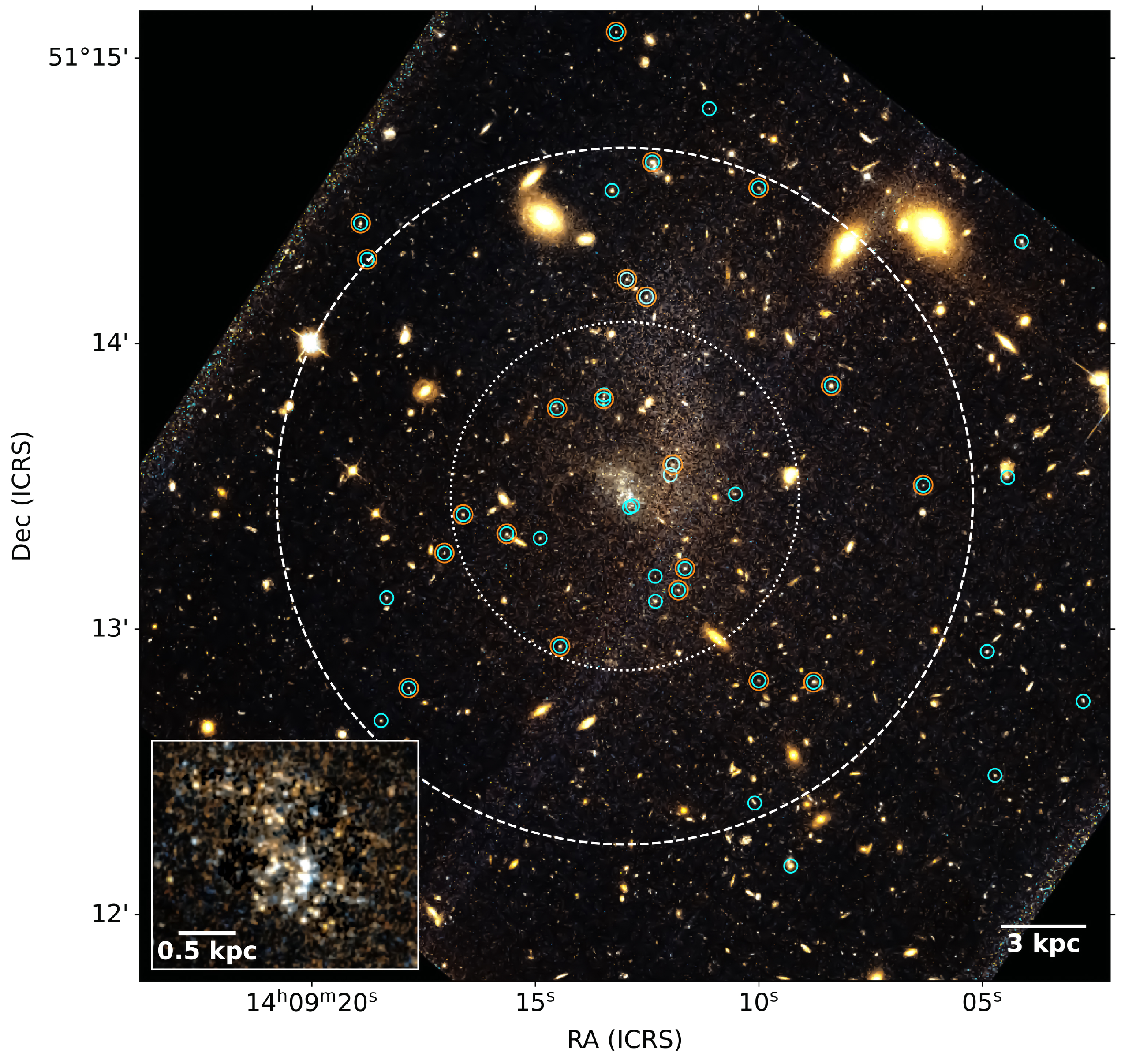}
    \caption{HST ACS/WFC F555W/F814W composite color image of UGC~9050-Dw1, which consists of very diffuse stellar light and an amorphous blue central clump of a relatively higher surface brightness. The inset at the lower left depicts a high-contrast zoom in of the central region of the UDG. The image is oriented such that north is up and east is left. The white dotted circular aperture marks $r_{\rm{UDG}}$ (36.6''), the best measured photometric boundary of the UDG, and the dashed circular aperture mark $2\times r_{\rm{UDG}}$ (73.2'') which is our final GC search radius. Cyan apertures with a radius of 35 pixels enclose all GC candidates within the image, with a total of 30 within the search radius. We encircle GCCs that fall within a narrow color cut ($0.8 < (V-I) < 0.96$) with a larger 50 pixel radius orange circle. GCCs of UGC~9050-Dw1 are only selected from those within $2\times r_{\rm{UDG}}$. GCCs outside of this radius are completely discounted since association is harder to discern photometrically at large radii.}
    \label{fig:HST_img}
\end{figure*}

\autoref{fig:HST_img} shows an RGB color composite constructed with the stacked F555W and F814W images. A blue, higher surface brightness clump at the presumed center of the UDG is apparent, for which we provide a high-contrast zoom-in  (see \autoref{subsec:optical_props} for further discussion of sources marked in the image).

The HST images used are the standard data products from the STScI archive. The \textsc{Dolphot v2.0} \citep{dolphin2000,dolphin2016} software was used for our photometric analysis. \textsc{Dolphot} utilizes the non-drizzled images that are the calibrated, flat-fielded individual exposures (flt.fits files). As a reference frame \textsc{Dolphot} uses the HST pipeline-provided calibrated, geometrically corrected, dither-combined image created by AstroDrizzle that includes a CTE correction (drc.fits files). We run \textsc{Dolphot} with the standard ACS and WFC3 parameters as provided in the user manual to align the individual HST exposures and then generate a combined point source catalog for each field. $V$- and $I$-band magnitudes for sources determined by \textsc{Dolphot} are derived via the HST to Johnsons-Cousins magnitude conversion factors presented in \citet{sirianni2005}. We then correct the derived $V$- and $I$-band magnitude quantities for Galactic extinction using the NASA/IPAC\footnote{\url{https://irsa.ipac.caltech.edu/applications/DUST/}} online tool, with values derived from the \citet{schlafly2011} extinction coefficients. Magnitudes presented in this work are Milky Way extinction corrected unless otherwise indicated.

Our point source completeness limits are determined via artificial star tests. Using the tools provided by \textsc{Dolphot}, we place nearly $100,000$ artificial stars into the ACS field of view. These artificial stars span a large color range from $-$1 to 2 in F555W$-$F814W (well beyond the range used to select GCs). We then measure the fraction of those stars that we recover as a function of apparent magnitude. We find we are $90\%$ complete to $m_{F814W} = 26.8$ and $50\%$ complete to $m_{F814W} = 27.4$. 



\subsection{VLA Observations}
\label{subsec:vla_obs}

UGC~9050-Dw1 was observed in the VLA D-configuration during July 2022, as a part of project 22A-225 (PI: D.~Sand). The total on-source integration time was approximately 4.3~hr. The data were simultaneously recorded with two correlator configurations, one with a 4~MHz bandwidth (approximately centered on the radial velocity of UGC~9050) and a channel width of 3.91~kHz ($\sim$0.8~\kms), and the other with a 32~MHz bandwidth and 62.5~kHz channels. The latter was in case the UDG was a foreground or background object, not at the same redshift as UGC~9050. For the remainder of this work we will only consider the former setup. The results of these observations are presented in the left panel of \autoref{fig:radio_img} and further discussed in \autoref{subsec:hi_props}.

The data reduction relied on the \hi\ pipeline\footnote{\url{https://github.com/AMIGA-IAA/hcg_hi_pipeline}} of \citet{jones2023b}. We refer the reader to that work for a full description, but we describe it briefly here. The pipeline began with a combination of manual  and automated flagging, then proceeded with gain and phase calibration using standard \texttt{CASA} tasks (Common Astronomy Software Applications; \citealt{mcmullin2007}). Overall, the data suffered from moderate radio frequency interference and about 20\% of visibilities were flagged. Imaging used automated masking and a Briggs robust parameter of 0.5 as a compromise between resolution and sensitivity. The spectral resolution was also smoothed to 5~\kms. Multiscale CLEANing was performed down to approximately 2.5$\sigma_\mathrm{rms}$. The RMS noise in the final image cube is 0.9~mJy/beam and the synthesized beam size is 45.2\arcsec$\times$48.5\arcsec. 

\subsection{Ancilliary Data}

\subsubsection{CFHTLS Observations}

We also use data from the Wide portion of the CFHTLS. This is a survey that was conducted between 2003 and 2009 and covers 171 square degrees in the $u$, $g$, $r$, $i$, and $z$ bands. UGC~9050 and UGC~9050-Dw1 are found in the  W3$-$1$-$3 and W3$-$2$-$3 fields, using the nomenclature presented in figure 4 of \citet{gwyn12}. The exposure time for the $g$-band stacks used in this work was 2500s, with 2375s in $r$, and 6150s in $i$, all with a pixel scale of 0.186 arcsec per pixel. The fields were downloaded directly from the Canadian Astronomy Data Centre (CADC) and have been processed by the Terapix 7 pipeline \citep{bertin2002}. The point-spread functions (PSFs) for those image stacks were also downloaded from the CADC, which were used for measuring dwarf's structural parameters. The construction and calibration of these utilized the MegaPipe data pipeline \citep{gwyn08} and is described in detail by \citet{gwyn12}. We provide a composite color image constructed from the $g$, $r$, and $i$ bands of the W3$-$1$-$3 field in the upper left panel of \autoref{fig:cfht_galex_img}, and a high-contrast image constructed from the $g$-band of the same field in the upper right panel of \autoref{fig:cfht_galex_img} and the background of \autoref{fig:radio_img}.

The CFHTLS images have the advantage of increased sensitivity to extended low surface brightness emission compared to HST ACS/WFC. The extended emission of UGC~9050-Dw1 also occupies a large fraction of the ACS field, which complicates background estimation. Therefore we supplement our HST data with the CFHTLS data to derive the magnitudes, colors, surface brightness, and radius for UCG~9050-Dw1.


\subsubsection{GALEX Observations}

Data from GALEX \citep{galex2005} were used to measure the star formation rate (SFR; see \autoref{subsec:optical_props}) of UGC~9050-Dw1 and UGC~9050. Both targets were observed in the NUV for $\sim$1600 s as part of the guest investigator program (GI5-028, PI: Balogh). This program did not include FUV observations and neither UGC~9050-Dw1 nor UGC~9050 were observed as part of the GALEX All-Sky Imaging Survey, therefore we are limited to NUV observations. We provide the GALEX image of UGC~9050-Dw1 in the bottom panel of \autoref{fig:cfht_galex_img}.

\subsubsection{Apertif Observations}
\begin{figure*}
    \centering
    \includegraphics[width=1.02\columnwidth]{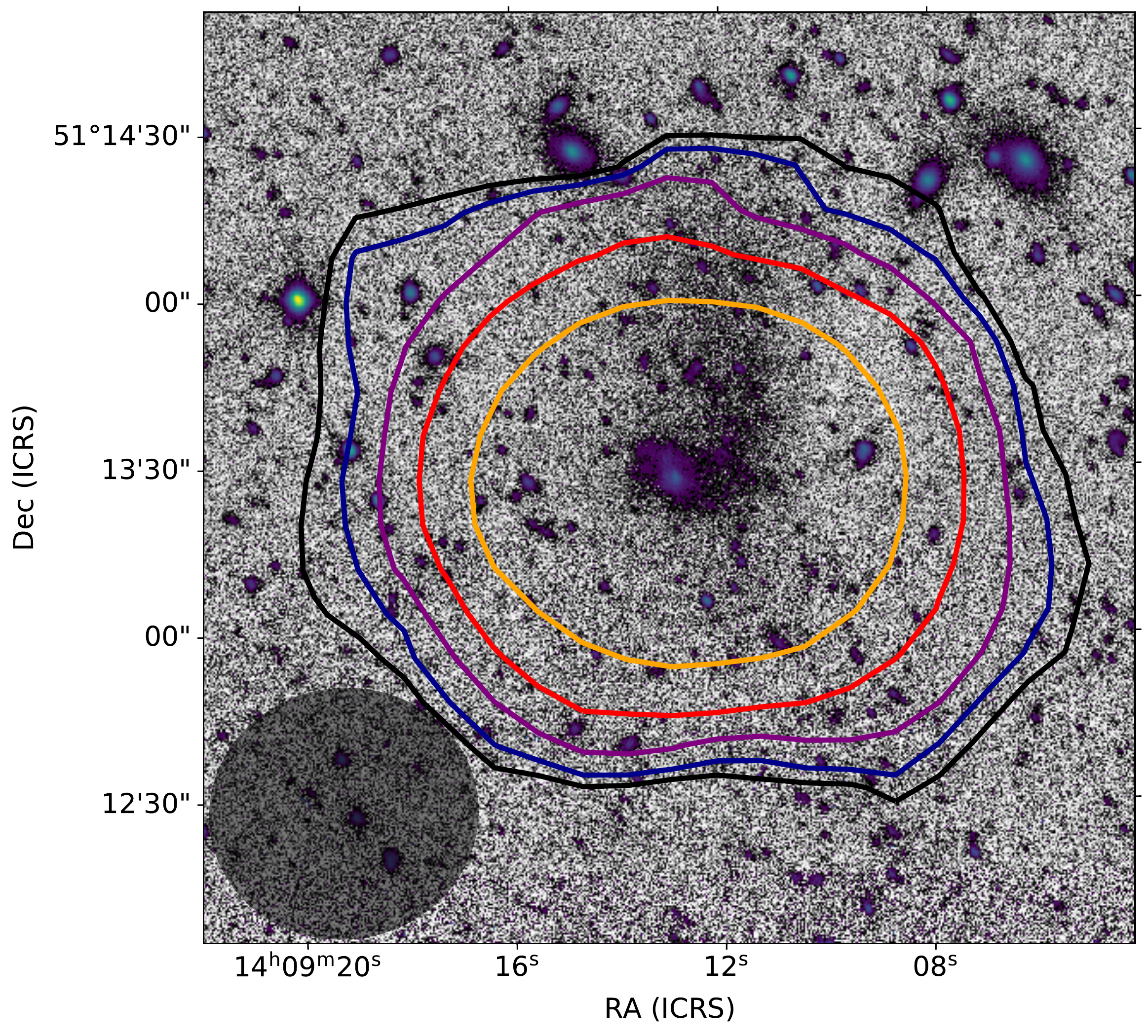}
    \includegraphics[width=1.02\columnwidth]{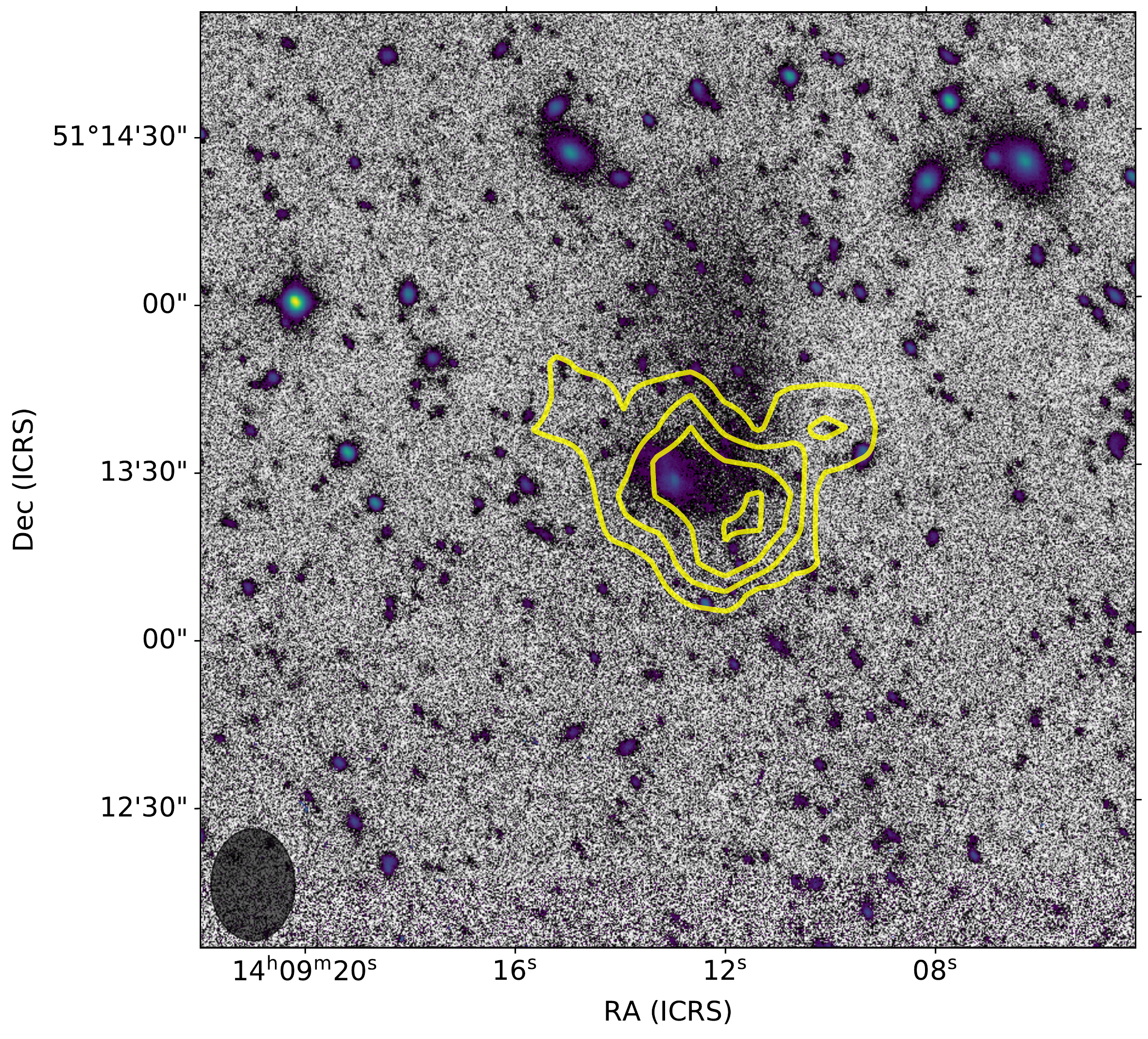}
\caption{\textit{Left}: 
A high-contrast $g$-band image of UGC~9050-Dw1 from the CFHTLS (see \autoref{fig:cfht_galex_img} for details) with contours of integrated \hi \ emission (VLA) overlaid. The outermost contour is 3$\sigma$ ($2.7\times10^{19} \; \mathrm{cm^{-2}}$ over 20~\kms) and each subsequent contour is double the previous one. The synthesized beam is shown as a semitransparent ellipse in the lower left corner. \textit{Right}: Same as left but with contours of integrated \hi \ from Apertif DR1 imaging shown in yellow, with the approximate Apertif beam shown in the lower left corner. The smallest central contour represents a peak with the highest \hi\ column density. However, we do not have physical contour levels because the data have not been cleaned. UGC~9050 lies almost directly to the right of this frame (slightly to the south).}
\label{fig:radio_img}
\end{figure*}

\begin{figure*}
    \centering
    \includegraphics[width=\columnwidth]{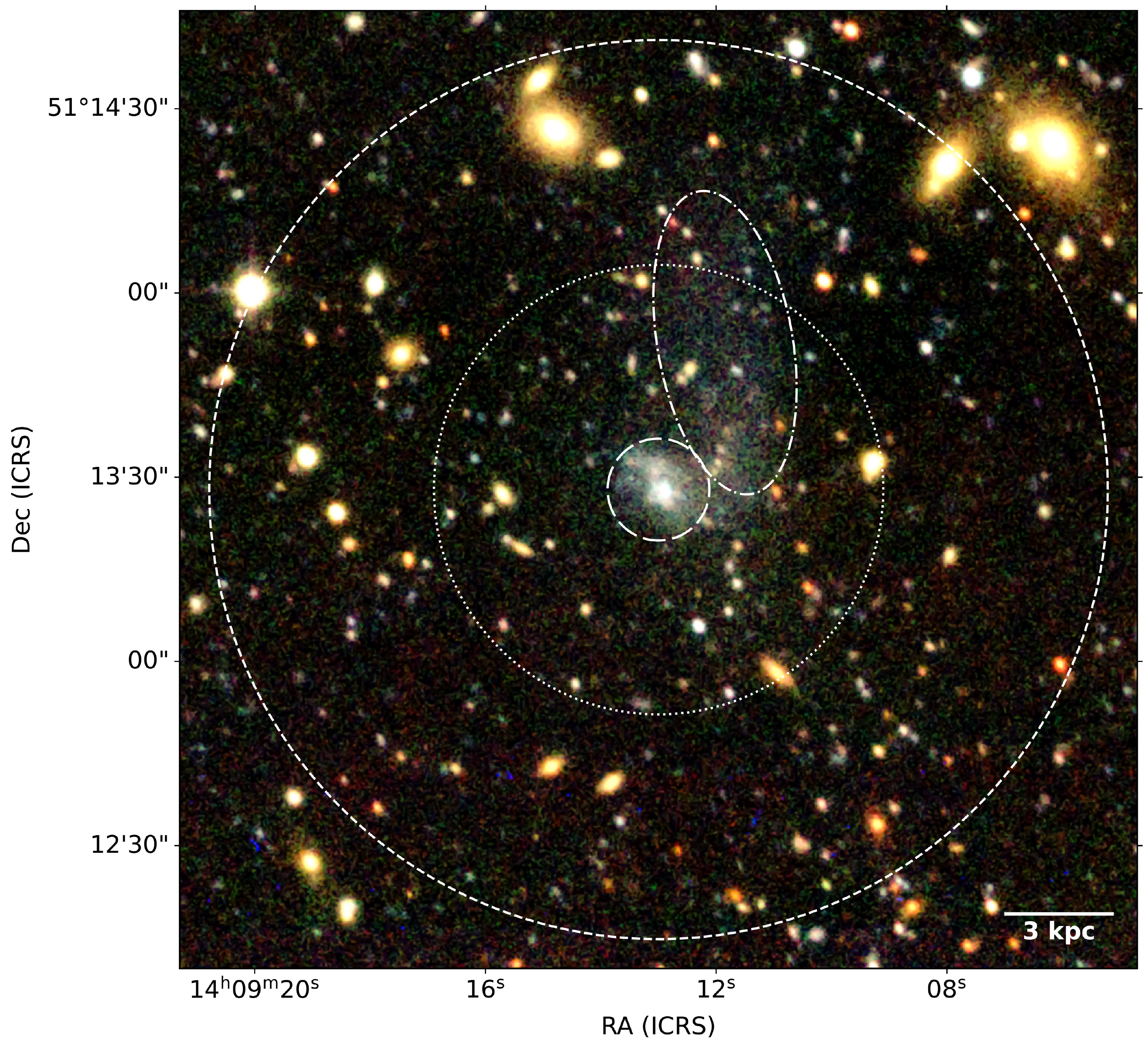}
    \includegraphics[width=\columnwidth]{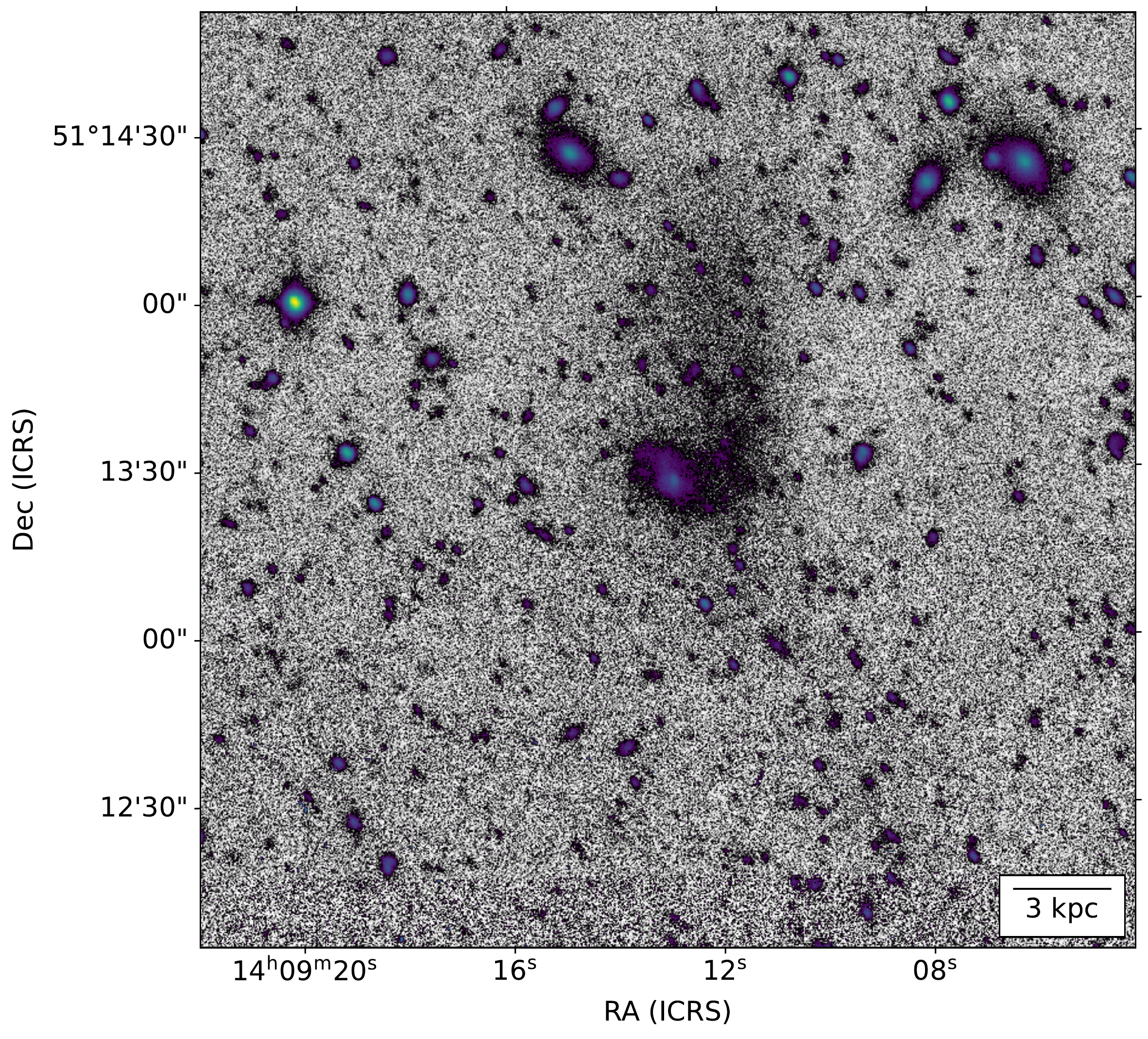}
    \includegraphics[width=\columnwidth]{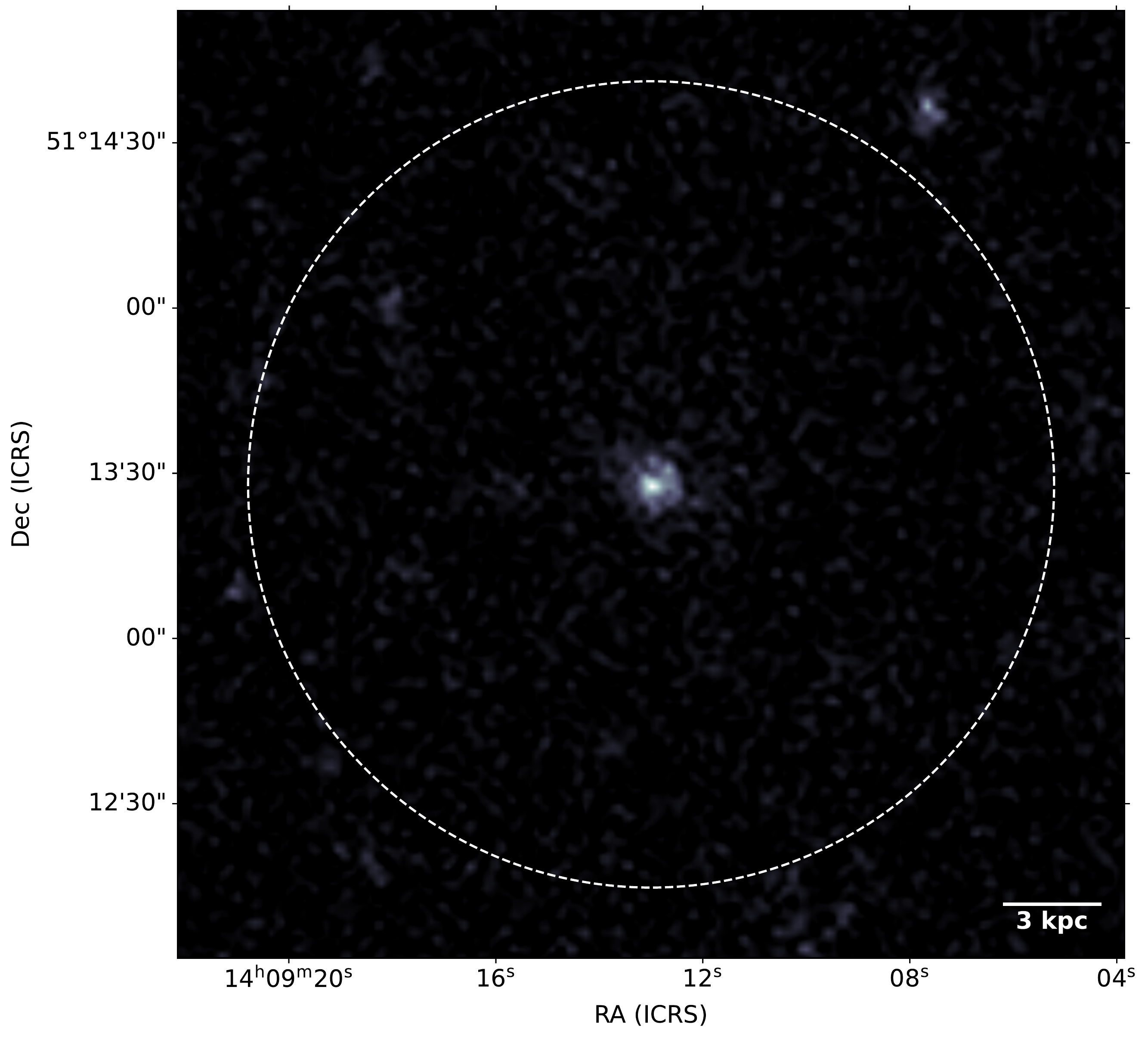}
    \caption{\textit{Upper Left:} CFHT composite color image of UGC~9050-Dw1, constructed from the $g$, $r$, and $i$-bands of the W3$-$1$-$3 field. As before the white dotted circular aperture marks $r_{\rm{UDG}}$ (36.6''), and the dashed circular aperture marks $2\times r_{\rm{UDG}}$ (73.2''). In this panel we also include a long-dashed circular aperture corresponding to the half light radius of the core and a dash-dotted elliptical aperture marking the tail.  \textit{Upper Right:} A high contrast $g$-band CFHT image of UGC~9050-Dw1 constructed with a histogram stretch, with a colored log-stretched mask to emphasize brighter regions. 
    \textit{Bottom:} GALEX $NUV$ image of UGC~9050-Dw1. UV emission in this UDG is associated with the central core feature. Additional speckles within the region of the UDG are consistent with noise.}
    \label{fig:cfht_galex_img}
\end{figure*}

We supplement our VLA observations with publicly available \hi\ imaging from the Apertif (Aperture tile in focus) imaging survey on the Westerbork Synthesis Radio Telescope (\citealt{adams2022}; seen in the right panel of \autoref{fig:radio_img}).\ 
An advantage of the Apertif imaging is its finer spatial resolution, which at the declination of UGC~9050-Dw1, is approximately $\sim15''\times20''$. These publicly available data\footnote{We obtained the cube from the ASTRON VO service: \url{https://vo.astron.nl/apertif_dr1/q/apertif_dr1_spectral_cubes/form}} have not yet been CLEANed or primary-beam corrected. We instead use these data as a qualitative check on the general morphology of UGC~9050-Dw1. We have compared the results from the dirty cube with upcoming cleaned and mosaiced Apertif data products (K.\ Hess, priv. comm.) and discuss this further in \autoref{subsec:hi_props}.


\section{The Physical Properties of UGC~9050-Dw1}
\label{sec:properties}

The morphology of UGC~9050-Dw1 is elongated and clumpy, indicative of a disturbance. It appears to have a more distinct plume in the northern direction with less structured diffuse emission in the south and around a bright clump. Specifically, the extended northern plume appears morphologically comparable to the tidal tails or plumes observed in a variety of interacting systems. Due to the complex morphology of the system, we will expound on the difficulty in deriving an exact size or center for this object. However, regardless of choice, the derived surface brightness and size discussed in the following fall well within the standard definition of a UDG (half-light radius $>1.5$ kpc; $g$-band surface brightness $>24$ mag arcsec$^{-2}$) -- criteria that the bright clumpy region itself nearly meets. 

In \autoref{fig:cfht_galex_img} it is evident that the bright NUV emission overlaps with the brightest region in the UDG within the high-contrast CFHTLS image. As seen in the HST image (\autoref{fig:HST_img}) this bright region is clumpy and displays a somewhat distorted appearance, preferentially in the east-west direction towards the start of the extended emission. The Apertif data (\autoref{fig:radio_img}) is distorted in the same east-west direction. The complex morphology of UGC~9050-Dw1 points to a dwarf-dwarf merger remnant or other strong interaction \citep[see e.g.,][]{paudel2018,kado-fong2020}. 

We note that it is unlikely that the central clump is a background galaxy, given the lack of clear indicative features, such as spiral arms or a redder bulge. In fact, in the radio data the \hi\ is also centered around the clumpy central region of the UDG and we find the velocity to be the same as the gas observed in UGC~9050.

\subsection{HI Content}
\label{subsec:hi_props}

To produce the VLA moment zero map (left panel, \autoref{fig:radio_img}) we used \texttt{SoFiA} \citep{serra2015} to create a source mask. We used the standard smooth and clip algorithm with no smoothing as well as Gaussian smoothing kernels approximately 0.5 and 1.0 times the beam size, plus boxcar spectral smoothing over 0 and 3 channels. The clip threshold was set to 3.5$\sigma$ and we required a 95\% reliability threshold. This returned just two objects, UGC~9050-Dw1 and UGC~9050 (the complementary moment zero map for UGC~9050 is shown in \autoref{sec:appendix}, \autoref{fig:host_radio_img}). The integrated \hi\ fluxes (corrected for the primary beam response) of each are $0.95 \pm 0.03$ and $3.02 \pm 0.05$~Jy~\kms, which at 35.2~Mpc equate to \hi\ masses of $\log M_\mathrm{HI}/M_\odot = 8.44\pm0.04$ and $8.94\pm0.04$, respectively, which are strikingly similar. The error represents a 10\% uncertainty estimate for the absolute calibration, which is the dominant source of error other than the distance. The \hi\ mass of UGC~9050-Dw1 is included in \autoref{tab:udg} in addition to optical and UV properties described in the following subsection. The documented heliocentric velocity of UGC~9050 is measured at $2001\pm5$~km/s. From the VLA data we derive a velocity of $1952.1 \pm 0.3$~km/s for UGC~9050-Dw1 via a Gaussian fit to the spectral line in \hi\ . This further supports the association of the two objects with each other.


We also estimate the size of the \hi\ distribution from the \hi\ mass-size relation derived in \citet{wang2016}, which has been shown to be tightly correlated and even holds for interacting galaxies. Assuming that the \hi\ in UGC~9050-Dw1 is in a disk, the \hi\ mass-size relation yields a \hi\ disk diameter $D_{\hi} = 9.5$ kpc. At the adopted distance of UGC~9050-Dw1 this corresponds to an angular size of $\approx56''$ in diameter, which is comparable to the resolution for the D-configuration ($48.5''$); the UDG is essentially unresolved. Therefore, while the VLA \hi\ morphology contains little evidence of disturbance in UGC~9050-Dw1, the lack of resolution makes this unconstraining. In the VLA data of UGC~9050, which is marginally resolved, there \textit{is} a suggestion of a disturbance to the \hi\ in the direction of UGC~9050-Dw1 (see \autoref{sec:appendix}). It is not certain that an interaction with the UDG is the cause, but it is plausible.

In the right panel of \autoref{fig:radio_img}, we also include higher spatial resolution \hi\ data from the first Apertif Data Release (DR1, \citealt{adams2022}). 
We follow a similar source finding procedure as described above using \texttt{SoFiA} and show the resulting moment zero map as contours on top of the same high contrast CFHTLS optical image of UGC~9050-Dw1. Since we do not clean or primary-beam correct these data (which is not feasible with the DR1 data), these contours do not have \hi \ column density values as in the left panel. However, we do check the general morphology with upcoming cleaned Apertif imaging and find that it is consistent with the results presented here (K.\ Hess, priv. comm.). There is a clear concentration of \hi\ near the central, UV-bright region of UGC~9050-Dw1 with hints of elongated morphology perpendicular to the northern diffuse extension, roughly in the direction of UGC~9050 (in the lower-right direction in the images). The deeper Apertif data that are forthcoming will provide more detail regarding the morphology.

\begin{table*}[]
\centering
\caption{UGC~9050-Dw1 Properties}
    \begin{tabular}{r|l|l|l|l}
    \hline\hline
         Property & Total & Core & Diffuse & Tail \\ \hline
         R.A. (J2000) & 14:09:13 & & & 14:09:11.847 \\
         Decl. (J2000) & +51:13:28 & & & +51:13:51.921 \\
         $D$ (Mpc) & $35.2\pm 2.5$ & & & \\
         $v_{\rm{helio}}$ (km/s) & $1952.1\pm0.3$ & & & \\
         $M_{V}$ & $-14.0\pm0.2$ & $-13.3\pm0.1$ & $-13.0\pm0.2$ & $-12.8\pm0.2$ \\
         $\log{\rm{L_{V}}/\rm{L}_{\odot}}$ & $7.5\pm0.1$ & $7.2\pm0.1$ & $7.1\pm0.1$ & $7.0\pm0.1$ \\
         $(V-I)$ & $0.8\pm0.3$ & $0.4\pm0.2$ & $1.0\pm0.3$ & $0.6\pm0.4$ \\
         $m_{g}$ & $18.9\pm0.2$ & $19.2\pm0.1$ & $20.4\pm0.2$ & $20.0\pm0.1$ \\
         $(g-r)$ & $0.3\pm0.3$ & $-0.2\pm0.2$ & $0.9\pm0.4$ & $0.2\pm0.2$ \\
         $m_{NUV}$ & $19.9\pm0.2$ & $19.9\pm0.2$ & -- & -- \\
         $r$ (arcsecond) & $\sim36.6\pm6$ & $8.3\pm1.8$ & -- & \\
         $r$ (kpc) & $\sim6.2\pm1$ & $1.4\pm0.3$ & -- & \\
         $\mu(g)$ (mag arcsec$^{-2}$) & $28.7\pm0.4$ & $24.7\pm0.3$ & &\\
         $\log{M_{*}/M_{\odot}}$ & $\sim$7.5 & $\sim$6.6 & $\sim$7.4 & $\sim6.7$ \\
         $\log{M_{\hi\ }/M_{\odot}}$ & $8.44\pm0.04$ & & & \\
         Projected distance (kpc) & $68.9 \pm 3.5$ & & & \\
         SFR (\msun\ yr$^{-1}$ $\times10^{-3}$) & $5.14 \pm 0.95$ & $5.14 \pm 0.95$ & -- & -- \\
    \hline
    \end{tabular}\\ [4pt]
    Rows: 1) R.A. in sexagesimal hh:mm:ss. 2) Decl in dd:mm:ss. 3) D: assumed distance in Mpc 4) $v_{\rm{helio}}$: Heliocentric velocity. 5) $M_{V}$: $V$-band absolute Vega magnitude determined from CFHT data. 6) $\log{\rm{L_{V}}/\rm{L}_{\odot}}$: $V$-band luminosity. 7) $V-I$ color determined from CFHT data in Vega magnitudes. 
    8) $m_{g}$: $g$-band apparent AB magnitude, determined from CFHT data. 9) $g-r$ color determined from CFHT data, in AB magnitude. 10) $m_{NUV}$: $NUV$ apparent magnitude determined from GALEX data, in AB magnitudes. 11) \& 12) $r$: Approximate radius. 13) $\mu(g)$: $g$-band surface brightness in AB magnitudes. The quantity in the ``total'' column corresponds to the mean surface brightness within $r$ ($\mu(g,\rm{r})$) and the quantity in the ``core'' column corresponds to the central surface brightness ($\mu(g,0)$) as it is dominated by the clump. 14) $\log{M_{*}/M_{\odot}}$: Logarithm of stellar mass using the M/L derived from \citet{zhang2017}. 15) $\log{M_{\hi\ }/M_{\odot}}$: Logarithm of \hi\ mass. 16) Projected distance from UGC~9050. 17) SFR: Star formation rate. \\
    Measurements with ``--'' signify that that region of the UDG is not contributing to a given measurement. Note that the diffuse component includes part of the tail. Uncertainty estimates do not include the distance uncertainty. \\
    \label{tab:udg}
\end{table*}

\subsection{Optical/UV Properties}
\label{subsec:optical_props}

\autoref{tab:udg} includes the optical and UV properties of UGC~9050-Dw1, for which $NUV$, $g$, and $r$-band magnitudes are presented in the AB system, while $V$ and $I$-band magnitudes remain in the Vega system. Similarly derived properties for UGC~9050 are presented in \autoref{tab:host}.

Due to the unique morphology of UGC~9050-Dw1, we split some of the derived properties between the ``core'', the ``diffuse'' structure, the ``tail'', and the entire object -- a choice driven by the difficulty in modeling the UDG given its morphology. As a result, determining a radius or edge for this UDG is non-trivial. 

We start with GALFIT \citep{peng2002} to model the inner ``core'' of the UDG and determine its properties. The GALFIT determined core size corresponds to a half-light radius of $1.4\pm0.3$~kpc or $8.3\pm1.8$ arcsec. 

However, GALFIT was unable to model the full surrounding extended emission, even after multiple techniques to boost the signal. Instead we used standard aperture photometry to characterize the remainder of the galaxy. To determine an approximate size of the UDG, we centered a circular aperture on the dwarf (determined by the center of the GALFIT model of the core region) and steadily increased the size of the aperture by 1 pixel (0.186 arcsec) until the additional flux was explainable as entirely background emission. With this method we find a circularized radius of $36.6\pm6$ arcsec. We settled on a circular radius centered upon the ``core'' due to the extended diffuse emission that falls both to the north and south of the core, and the highest density \hi\ contours coincident with this region of the UDG (see \autoref{fig:radio_img}). An elliptical aperture would similarly be centered on the same region due to the location of the diffuse emission and \hi. Additionally the ellipticity in by-hand aperture photometry must be selected by eye, so we chose a circular ellipse for simplicity. We caution the reader that the final circularized radius of $36.6\pm6$ arcsec, which we forthwith refer to as $r_{\rm{UDG}}$, is merely an estimate and not a robust measured edge of the UDG. The properties listed in the ``Total'' column of \autoref{tab:udg} are calculated within $r_{\rm{UDG}}$. The errors on the derived photometric quantities are large, reflective of the uncertainty of the dwarf radius.

Derived photometry documented in the ``Diffuse'' column is simply the ``Core'' quantities subtracted from the ``Total'' quantities.

At $r_{\rm{UDG}}$ there remains excess diffuse emission to the north beyond 36.6 arcsec, associated with the ``tail'' feature of UGC~9050-Dw1 (see \autoref{fig:HST_img} and \autoref{fig:cfht_galex_img}; upper panels) that even aperture photometry fails to pick up. We instead opt to measure approximate photometry for this distinct tail region with a tailored elliptical aperture. We manually place an elliptical aperture centered on the approximate visible center of the tail (coordinates in \autoref{tab:udg}). The ellipse has a semi-major axis of 25 arcsec (4.2 kpc), a semi-minor axis of 11 arcsec (1.9 kpc), and a position angle of 10 degrees to the east. GALFIT was then used to determine the color and magnitudes of the properties within the tail, using the arbitrary ellipse as limits. The errors presented are likely underestimates as they do not factor in the original determination of the spatial extent of the tail, which was somewhat subjective. Note that the placement of the ellipse for the tail means that some of what we refer to as the diffuse emission region is included within this ellipse and we do no separate out the two. 

To quantify the uncertainties of the structural and photometric properties of UGC~9050-Dw1 in the CFHTLS images, we used the procedures from \cite{bennet2017} (see also \citealt{merritt2014,bennet2018,jones2021}). In brief, after the observational properties are derived we inject 100 simulated galaxies with the same observed properties as the various components of UGC~9050-Dw1 into the original CFHTLS image. Properties of the simulated dwarfs are fit in the same way we determine the properties of UGC~9050-Dw1. Then the scatter in the measurements of the simulated galaxies are used to characterize the uncertainty of the observational properties of UGC~9050-Dw1.

We also provide the NUV magnitude obtained from the GALEX survey \citep{galex2005} and use the relation from \citet{IglesiasParamo2006} to derive the star formation rate. The NUV flux is determined with aperture photometry using an aperture equivalent to two half-light radii for the core model ($16.6''$). We find that any additional NUV flux outside the core region is entirely consistent with noise, indicating that all of the UV emission comes from the central ``core'' of the UDG (see \autoref{fig:cfht_galex_img}, bottom panel). This was done by comparing the NUV flux from just the core to the NUV flux within the full circular radius (36.6''), which were equivalent after accounting for noise.

While we do include a derived stellar mass in \autoref{tab:udg} ($\sim 3.2\times10^{7}$~\msun\ for UGC~9050-Dw1, compared to $3.7\times10^{8}$~\msun\ for UGC~9050) using the relations from Table 1 of \citet{zhang2017}, this mass should be considered a guide and has significant uncertainty. Using the $(V-I)$ color and $M_{V}$ we derive a stellar mass-to-light ratio $\Gamma_{*}=0.91$ for the full UDG, $\Gamma_{*}=0.24$ for the core, $\Gamma_{*}=1.76$ for the total diffuse component, and $\Gamma_{*}=0.47$ for the tail alone. 

\section{Selection of Globular Cluster Candidates}
\label{sec:gcc_selection}

On average GCs have radii of just a few parsecs \citep{brodie2006}. If we assume a nominal radius of 5 pc, at the distance of UGC~9050-Dw1 (35 Mpc) this corresponds to a maximum angular diameter of $\sim 0.03''$. The FWHM of the HST PSF has been determined to be $\sim0.15''$ for ACS and  $\lesssim 0.1''$ for WFC3. Therefore, all GCs will appear as approximately point sources in our images. 

The selection of globular cluster candidates (GCCs) follows that described in \citet{jones2021}, which we briefly summarize here. Both HST filter images are processed through the \textsc{Dolphot} pipeline, which includes masking out known bad pixels, creating a sky map, aligning the images from each filter, and ultimately performing photometric analysis. \textsc{Dolphot} yields a total of 211,225 sources for our pointing. We then further limit this catalog to sources classified as either bright or faint stars in the \textsc{Dolphot} pipeline (187,891 sources) and having no flags in the photometry, with a signal-to-noise ratio (S/N) $> 5$ (3,415 sources). Cutting on objects identified as ``stars'' does not restrict to solely perfect point sources, and further cuts are required. Specifically, in both HST filters sharpness is required to fall within the range of $-$0.3 to 0.3 (restricting to 885 sources) and roundness is required to be less than 0.3 (further restricting to 674 sources). This aims to exclude extended or overly compact sources, and any that may be elongated. In either band we set a maximum limit on crowding to 0.5 mag, and require \textsc{Dolphot} magnitude uncertainties to be smaller than 0.3 mag (restricted to 524 sources). We experimented with adjusting the crowding parameter in order to identify more sources in the bright ``core'' region, expanding from 0.5 mag to 1 mag and 2 mag limits. However, no additional sources were ultimately identified as GCCs.  

A cut in concentration index is employed in addition to the above criteria for GC candidate selection. The concentration index is determined by comparing flux in concentric circular apertures of 4 and 8 pixel diameters on the background-subtracted F814W image. We use the same concentration definition as \citet{jones2021}, which aims at eliminating diffraction spikes, background galaxies, and other sources that would have high concentration indices (see similar approaches in e.g., \citealt{peng2011,beasley2016,saifollahi2021}). We define this concentration index as $C_{4-8} = -2.5\log_{10}{(\frac{N_{4\rm{pix}}}{N_{8\rm{pix}}})}$ where $N_{4\rm{pix}}$ represents the sum of flux values in a 4 pixel aperture. We allow the concentration index parameter to span the range $0.2-0.8$ mag. This largely excludes any remaining background galaxies that exhibit discernible structure and other image artifacts. This cut further restricts our source catalog to 228 sources. 

To select GCCs we also employ a color cut of $0.5 < (V-I) < 1.5$, within which globular clusters are expected to lie \citep[e.g.,][]{brodie2006}, which will largely eliminate blue star-forming clumps. Such a cut yields 127 sources. Lastly, in an effort to maximize the number of GCCs while minimizing the number of contaminants, we employ a cut in magnitude in addition to the cuts above. Similar to the procedure of \citet{peng2011}, we assume a Gaussian dwarf elliptical globular cluster luminosity function (GCLF) of \citet{miller2007}, which peaks at $M_{I}=-8.12$ with $\sigma_{M_{I}}=1.42$ mag. At the distance of UGC~9050-Dw1, this corresponds to a peak of $m_{I} = 24.61$. We use this peak to serve as a brightness minimum, instead of our completeness limit, as a means to minimize any possible stellar contaminants. Likewise, we also employ a brightness maximum above which any point sources are assumed to be foreground stars, $M_{I} > -12.38$ (or $3\sigma$ from the mean of the GCLF). Thus in apparent magnitude all UGC~9050-Dw1 GCCs were selected within the range $20.35 < m_{I} < 24.61$, yielding 42 sources in total. With this cut, we sample approximately half of the luminosity function by number. As a result, our final GCC counts will be completeness corrected by multiplying by 2. 

\begin{figure}
    \centering
    \includegraphics[width=\linewidth]{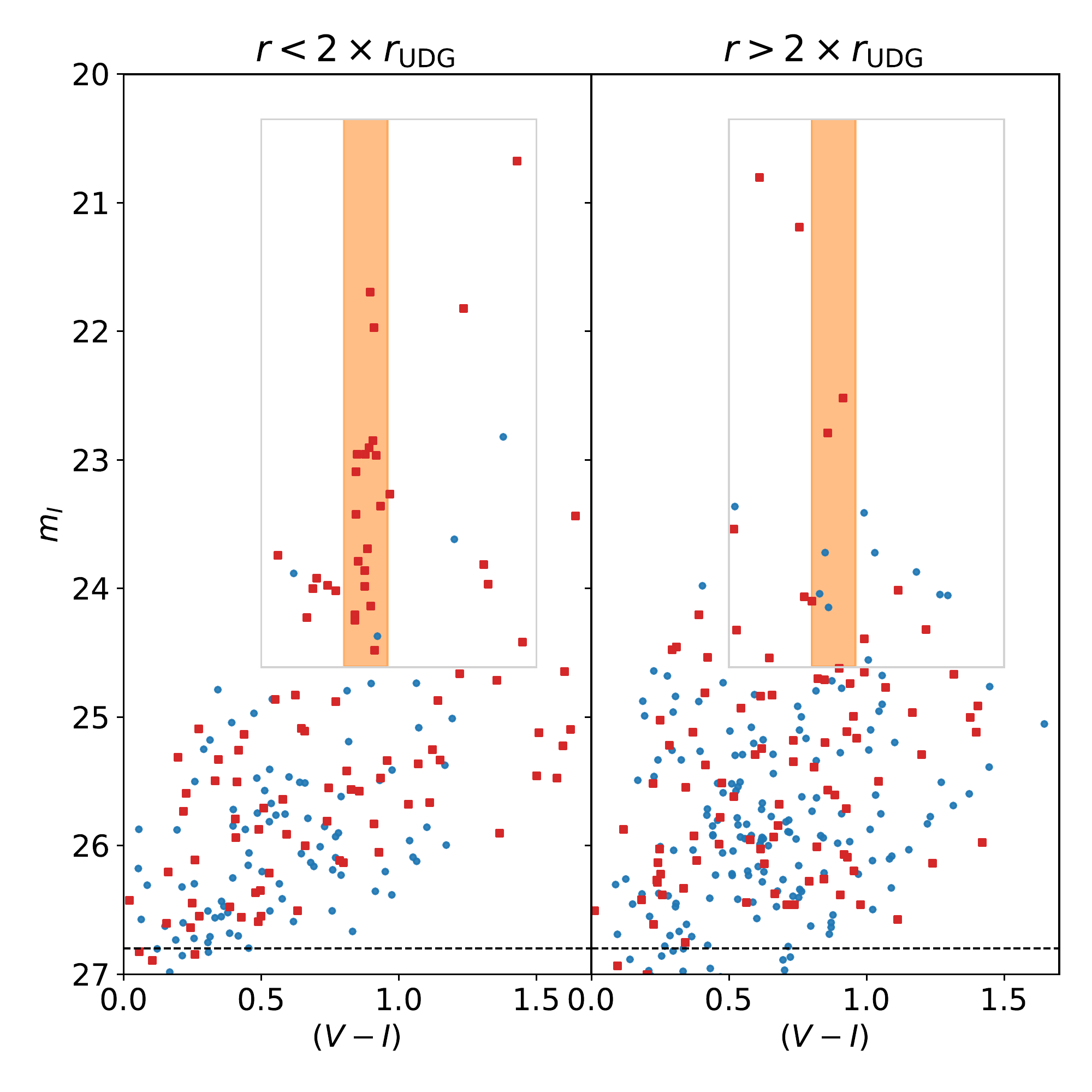}
    \caption{Color-magnitude  diagrams (CMDs) of point sources in the HST/ACS frame both inside (left panel) and outside (right panel) the UDG that pass the S/N, sharpness, roundness, and uncertainty cuts in \textsc{Dolphot} (blue circles). Sources that pass the subsequent concentration cut are plotted as red squares. The gray outlined box shows where the color and magnitude cuts for GCCs lie in the CMD space. The horizontal black dashed line marks the $90\%$ completeness limit. We highlight a monochromatic color region in orange ($0.8 < (V-I) < 0.96$) for comparison purposes, where the range corresponds to $\pm 1\sigma$ on the median of the blue GCCs. GCCs that also pass the concentration cut and fall within the gray box of this CMD that lie within this narrow color window are highlighted by orange circular apertures in \autoref{fig:HST_img}.}
    \label{fig:CMD}
\end{figure}

We include a color-magnitude diagram in \autoref{fig:CMD} of the acceptable \textsc{Dolphot} point sources (blue points) and the GC color and magnitude selection box in gray. Sources plotted as red squares are those that also pass the concentration cut described above. Point sources that fall within the selection box and pass the concentration cut (i.e., red points within the gray box) are circled in teal in \autoref{fig:HST_img}. The narrow color range highlighted in orange corresponds to the $\pm 1\sigma$ range on the median of the blue GC candidates (see \autoref{subsubsec:gc_color} for additional details). Point sources that fall within this narrow color range are subsequently circled in orange in \autoref{fig:HST_img}. 

\section{The GC System of UGC~9050-Dw1}
\label{sec:gc_props}


\subsection{Globular Cluster Abundance}
\label{subsec:gc_abundance}

After restricting the catalog of point-like sources to those with colors and concentration indices consistent with GCs, we ultimately consider GCs within $2\times r_{\rm{UDG}}$ or 73.2 arcsec. This choice is motivated by GCC searches in UDGs performed by other groups that search for globular clusters within 1.5, 2, 2.5, or 3 times the effective radius of the UDG \citep[see e.g.,][and many others]{jones2021,muller2021,danieli2022,janssens2022,jones2023}. Likewise, due to the complex morphology of the system we opt to measure everything within a large radius and subtract out contaminants rather than attempt to extrapolate the GC distribution. At the distance of UGC~9050-Dw1, $2\times r_{\rm{UDG}}$ corresponds to a physical radius of $\sim12.4$~kpc. Within this aperture we count a total of 30 GCCs. These candidates are marked with cyan circles in \autoref{fig:HST_img}, with the UDG radii marked by white dotted and dashed apertures ($r_{\rm{UDG}}$ and $2\times r_{\rm{UDG}}$). 

Compared to the ACS/WFC field-of-view (202x202''), UGC~9050-Dw1 fills a substantial fraction. Therefore to estimate the contaminant count rate for false GCCs we utilize the parallel WFC3 field. Within this field we use the same basic cuts for point-like sources in addition to the magnitude, color, and concentration index cuts described in \autoref{sec:gcc_selection} used to obtain the GCCs in the ACS field. In the WFC3 field a total of 7 sources pass these selection criteria. However, the WFC3 field-of-view (160''x160'') is smaller than the ACS field-of-view. To account for this we scale the number of the sources found in the parallel field relative to the ACS detector pixel area, yielding a total of 11 expected GC contaminants in the full ACS area. Finally, this quantity is then scaled down to the aperture area we use to search for GCCs ($2\times r_{\rm{UDG}}$), resulting in an estimated total of $4\pm2$ contaminants rounded to the nearest whole number.

While the parallel field is close to the pointing of the UDG (RA=14:09:45.89, Dec=+51:12:25.82), it may not be perfectly representative of the field that the UDG lies in, due to small number statistics. Therefore we perform a check of the foreground contaminant estimate with the accessible via web-form \textsc{TRILEGAL v1.6} \footnote{\url{http://stev.oapd.inaf.it/cgi-bin/trilegal_1.6}} simulation \citep{girardi2005,girardi2012}.
\textsc{TRILEGAL} (TRIdimensional modeL of thE GALaxy) is a Milky Way star counts model based upon population synthesis. \textsc{TRILEGAL} builds a geometric model of the Milky Way, with the thin disk, thick disk, halo, and bulge operating as the key primary components, and with each containing specific stellar populations. The actual geometric parameters of the Milky Way used in the model are calibrated using wide-area survey data. We use \textsc{TRILEGAL} to simulate a photometric catalog of stars, running the simulation with the default settings and a Chabrier log-normal initial mass function (IMF; \citealt{chabrier2001}). Extinction corrections are applied to the output photometry in the same manner as our HST observations (see \autoref{subsec:hst_obs}). Instead of querying a simulated area to just the ACS camera field-of-view we search a full square degree centered on the coordinates of UGC~9050-Dw1. All of the simulated stars in this field are then down-sampled randomly 10,000 times to both the area of the ACS field and the WFC3 field to ensure our contamination estimates are consistent. After applying the color/magnitude cuts consistent with GCs we find an average of $6\pm2$ contaminants in the ACS field, $4\pm2$ contaminants in the WFC3 field, and $2\pm1$ contaminants in the area of UGC~9050-Dw1. These numbers are smaller but consistent with the results obtained from the WFC3 parallel field described above, implying that we have obtained a reliable contaminant estimate. The numbers derived from WFC3 are likely higher since \textsc{TRILEGAL} only simulates Milky Way foreground contaminants and not other sources of background contaminants. To optimize our GC candidate selection, we will use contaminant counts derived from the WFC3 field ($4\pm2$ in the UDG area). 

With a contaminant estimate we can now derive a GC abundance for UGC~9050-Dw1. With a total of 30 GCCs within the aperture of the UDG we then subtract the contaminant counts, yielding 26 likely GCs in our sample. The error on this measurement consists of the Poisson error on our raw GCC counts ($\sigma_{P}=\sqrt{30}$) added in quadrature with the error on the contaminants ($\sigma_{bg}=2$), yielding a final error of $\sigma = \sqrt{34} \approxeq 6$. We correct the final result for covering only half of the GCLF by multiplying by 2, as discussed in \autoref{sec:gcc_selection}. This yields a GC abundance of $52\pm12$.

\begin{table}[]
\centering
\caption{UGC~9050-Dw1 GC Abundance}
    \begin{tabular}{r|l}
    \hline\hline
         Identified GC Candidate Abundance & 30 \\
         Identified Contaminant Abundance & $4\pm2$ \\
         Most Probable GC Abundance for half the GCLF & $26\pm6$ \\
         Most Probable Final GC Abundance & $52\pm12$ \\
    \hline
    \end{tabular}
    \label{tab:gc_counts}
\end{table}

\begin{figure*}
    \centering
    \includegraphics[width=0.75\textwidth]{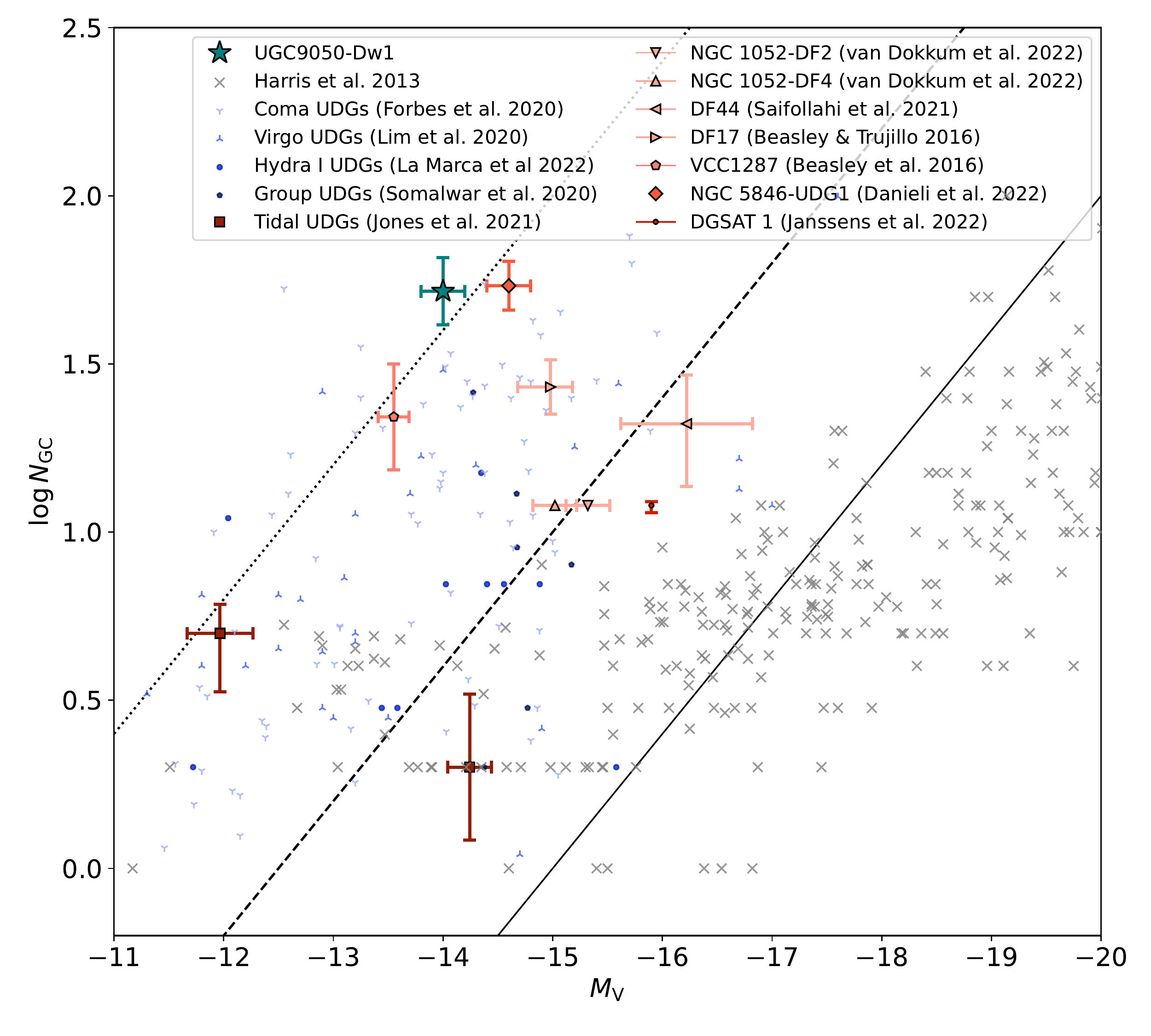}
    \caption{Counts of globular clusters ($N_{\rm{GC}}$) plotted as a function of $V$-band absolute magnitude ($M_{V}$). We compare to a number of samples in the literature, including a general galaxy sample, cluster UDGs, group UDGs, and other unique UDGs from a variety of environments. The solid, dashed, and dotted black lines mark specific frequency values of 1, 10, and 100 respectively (right to left). Among all of these UDG samples, UGC~9050-Dw1 displays an exceptionally rich GC abundance and high specific frequency.}

    \label{fig:ngc_mv}
\end{figure*}

The GC counts for this system are plotted in \autoref{fig:ngc_mv} along with a number of UDG samples and individual UDG studies. This literature sample includes nearby galaxies \citep[spanning ellipticals, spirals, lenticulars, and irregulars;][]{harris2013}, Coma cluster UDGs \citep{forbes2020}, Virgo cluster UDGs \citep{lim2020}, and Hydra I cluster UDGs \citep{lamarca2022,iodice2020}, and group UDGs compiled in \citet{somalwar2020}. We also compare to the tidally puffed-up UDGs studied in \citet{jones2021} and some well-studied UDGs in the literature: UDGs with monochromatic GC populations such as Dragonfly 2 (DF2) and Dragonfly 4 (DF4) in the NGC 1052 group with updated measurements from \citet{vandokkum2022}, NGC 5846-UDG1/MATLAS19 in the NGC 5846 group \citep{muller2021,danieli2022}, and DGSAT 1 in a low-density environment associated with the Pisces-Perseus supercluster \citep{janssens2022}. Magnitude errors for NGC 5846-UDG1 are not quoted in \citet{danieli2022} so we utilize those provided in \citet{muller2021}. Last, we include UDGs with rich GC systems - updated measurements of Dragonfly 44 (DF44; a Coma UDG) recorded in \citet{saifollahi2021}, Dragonfly 17 (DF17; a Coma UDG studied in \citealt{beasley2016}), and VCC 1287 (a Virgo UDG studied in \citealt{beasley2016A}). Magnitude errors are not quoted for the Dragonfly objects so we adopt the $g$-band errors from \citet{vandokkum2015} and assume they are the same for $V$-band. We convert the VCC 1287 $u$ and $g$-band photometry documented in \citet{pandya2018} to $V$ using documented SDSS conversions\footnote{\url{http://classic.sdss.org/dr4/algorithms/sdssUBVRITransform.html}}.

In contrast to even the most GC-rich UDGs in the literature and the tidally puffed-up UDG sample of \citet{jones2021}, UGC~9050-Dw1 contains an exceptional abundance of globular clusters. Considering objects just within 2$\sigma$ of the estimated magnitude and GC abundance of UGC~9050-Dw1, only two Coma cluster UDGs and updated measurements of NGC~5846-UDG1 \citep{danieli2022} compare. 


\subsection{Properties of the Globular Cluster Population}
\label{subsec:gc_props}

\subsubsection{The Radial Profile}
\label{subsubsec:rad}

\begin{figure}
    \centering
    \includegraphics[width=\linewidth]{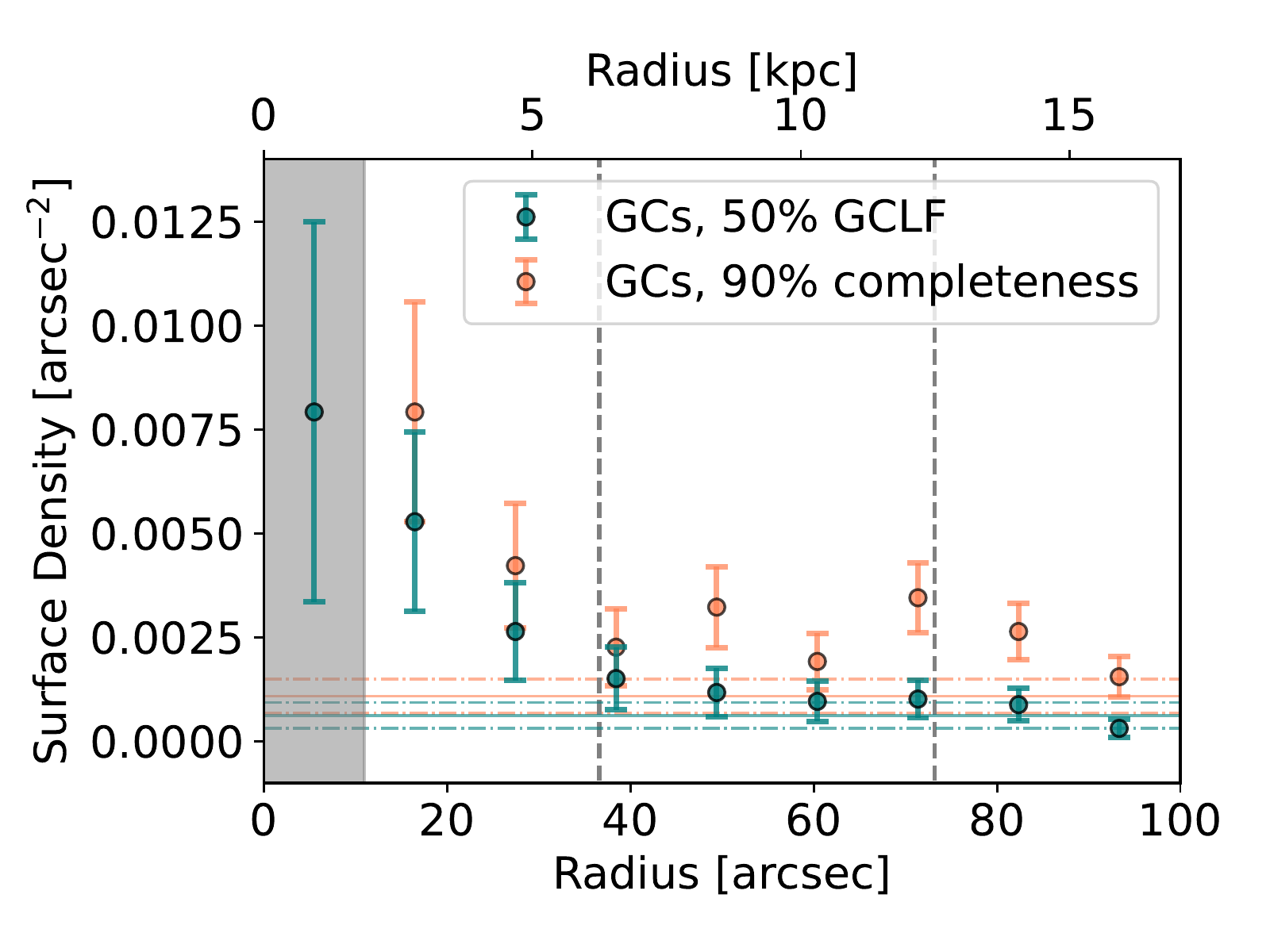}
    \caption{GC surface density in concentric annuli $0.3\times r_{\rm{UDG}}$ wide. In teal we present our 30 GCCs with corresponding Poisson errors. In orange we present GCCs if we allow for fainter GCs down to our $90\%$ completion limit. The vertical gray lines indicate $r_{\rm{UDG}}$ and $2\times r_{\rm{UDG}}$. The horizontal solid lines indicate the parallel WFC3 field contaminant density with dashed-dotted lines to mark the corresponding Poisson error. The gray region indicates the center of the UDG ($<11''$) which contains the bright central clump. We do not show the central orange point for readability, where 9 GCCs lie, which is equivalent to $y=0.02$. The GC radial distribution approximately follows the expected distribution shape within the scatter due to small bin counts, with a clear central concentration and exponential decrease to the background level at $r\gtrsim 2\times r_{\rm{UDG}}$.}
    \label{fig:radial}
\end{figure}

Here we consider the projected spatial distribution of the GC population of UGC~9050-Dw1. As is evident in \autoref{fig:HST_img} there appears to be a relatively central concentration of globular clusters. To quantify this further, we count the number of GCs per area in annuli of increasing multiples of $0.3r_{\rm{UDG}}$ from the center. These results are presented in \autoref{fig:radial} in teal. We include Poisson errors on our GC counts within each annulus bin. The expected contaminant counts per unit area within $2\times r_{\rm{UDG}}$ by a horizontal solid line with respective Poisson errors marked by a dashed-dotted line. By $\sim 2\times r_{\rm{UDG}}$ the GC counts are consistent with the background level, while the majority of the GCs project within $r_{\rm{UDG}}$. The gray shaded region marks the central $11''$ (1.9~kpc) of the UDG that contains the starry clump within our bin spacing. Because the center is very crowded it is more challenging to detect GCs reliably.

The orange points in \autoref{fig:radial} are for GCCs in UGC~9050-Dw1 if we extend the $I$-band magnitude cut discussed in \autoref{sec:gcc_selection} from $M_{I} < -8.12$ down to our $90\%$ completeness limit, which corresponds to $M_{I} < -5.96$ or 2 magnitudes fainter than the peak of the GCLF. For both cases we maintain the requirement that $M_{I} > -12.38$. Going down to fainter magnitudes we obtain 61 GCCs with a corresponding 7 contaminants within $2\times r_{\rm{UDG}}$ determined from the parallel WFC3 field. This relaxed cut may allow for more foreground contaminant stars, but we use this broader magnitude range simply as a consistency check. We note that with this extended cut, 7 GCCs fall within the gray shaded region which is well beyond the $y$-axis of \autoref{fig:radial} ($y = 0.02$).



On average, the spatial distribution of GC populations is expected to be cored with an increasingly steep outer slope \citep[e.g.,][]{brodie2006}. The general trend of our GC surface density displays such a distribution. The GC radial distribution for candidates down to the $90\%$ completeness limit approximately follows that of our more conservative magnitude cut.

\subsubsection{Globular Cluster Colors}
\label{subsubsec:gc_color}
\begin{figure}
    \centering
    \includegraphics[width=\linewidth]{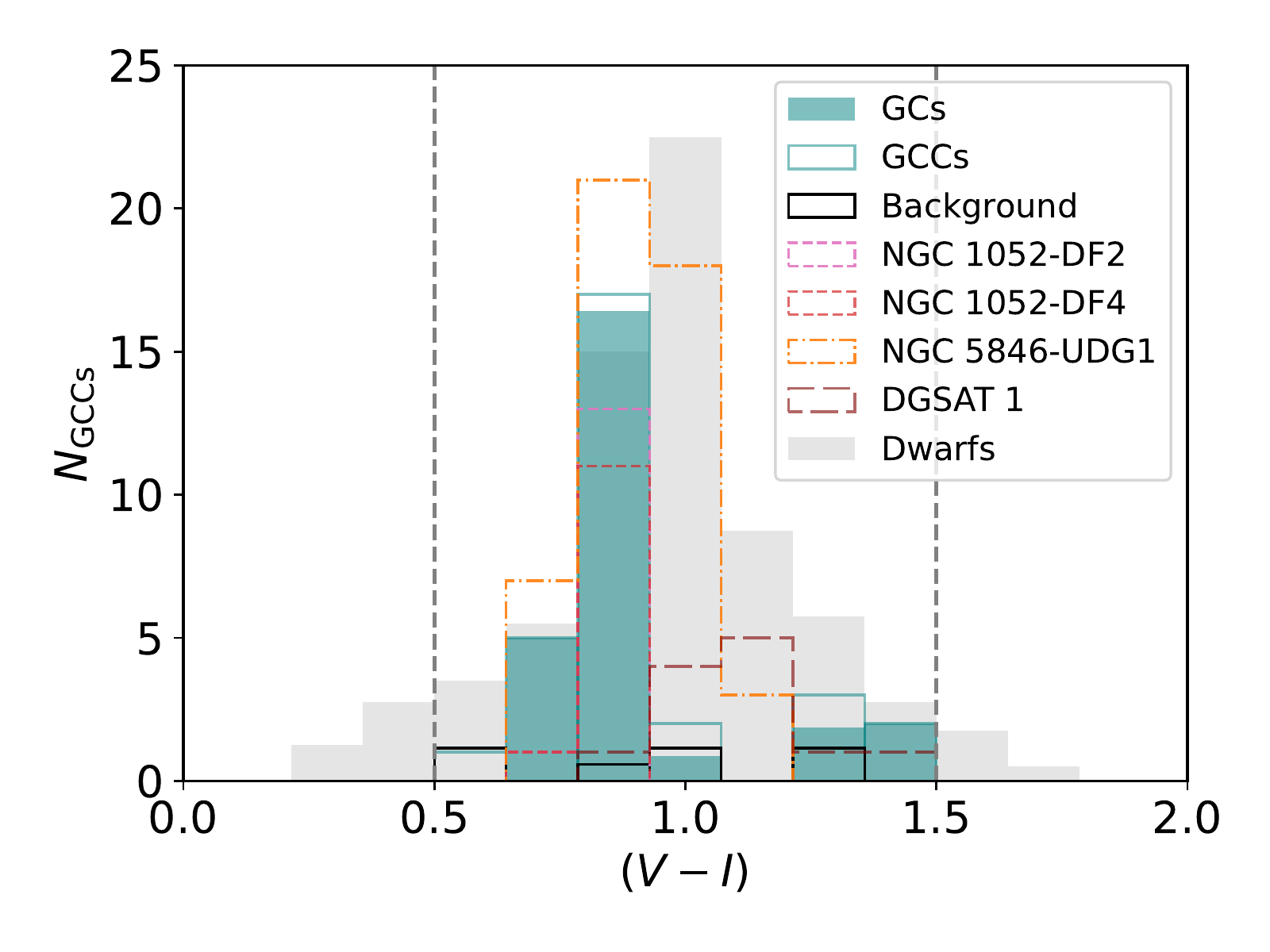}
    \caption{$(V-I)$ color distribution of UGC~9050-Dw1  plotted in solid teal bars. The outlined teal bars are the counts before subtraction of the contamination measured in the parallel WFC3 field, which is plotted by the solid black bars. The background is normalized to sum to $4$ for the expected contaminants within the UDG aperture. We plot other UDGs containing monochromatic GC populations and in the background we plot a sample of general dwarf galaxies \citep{sharina2005,georgiev2009} as gray bars, arbitrarily normalized in order to span the same vertical range as the small UDG sample. The vertical dashed lines mark the color cut for our GCCs. The colors of our GCCs follow a similar distribution to NGC 5846-UDG1, but lean more to the blue, like NGC 1052-DF2 and DF4. }
    \label{fig:color_dist}
\end{figure}

The color distribution of the 30 GCCs in UGC~9050-Dw1 is relatively blue and monochromatic. We present the $(V-I)$ color distribution of our GCCs in \autoref{fig:color_dist} with solid teal bars, in addition to other UDGs in the literature that have also been found to have monochromatic GC populations. The gray bars in the background constitute the GC color distribution of the average dwarf galaxy population, compiled from \citet{sharina2005} and \citet{georgiev2009}. It is evident that the GCC colors of UCG~9050-Dw1 are neither uniformly nor normally distributed, as one might expect based on the general dwarf galaxy GC color distribution. Across the full GCC color range we find a mean $(V-I) = 0.89$ mag, median $(V-I) = 0.86$ mag, and standard deviation $0.19$ mag for UGC~9050-Dw1 whereas the dwarf galaxy population exhibits median $(V-I)= 0.96$ and $\sigma_{(V-I)}=0.26$. While a majority of the GCCs in UGC~9050-Dw1 lie on the blue side of color space ($83\%$), 5 have $(V-I) > 1.24$ ($\frac{1}{6}$ of the GCCs). Two of the contaminants found in the WFC3 parallel field overlap with these red sources, indicating that it is likely at least several of these red GCCs are contaminants (if not all). 
If we only consider the blue GCC population between $0.6 \leq (V-I) \leq 1.2$ we find a mean $(V-I) = 0.85$, median $(V-I) = 0.88$, and $\sigma_{(V-I)}=0.08$; a substantially tighter spread than the dwarf galaxy population, and analogous to UDGs with monochromatic GCs.

So-called monochromatic GC populations have been observed in several other UDGs (NGC~1052-DF2 and NGC~1050-DF4 from \citealt{vandokkum2022}; NGC~5846-UDG1 from \citealt{muller2021,danieli2022}; DGSAT 1 from \citealt{janssens2022}). One proposed explanation suggests these GC populations formed in a single starbust event, yielding GCs of the same color and metallicity \citep{lee2021,vandokkum2022}. The incredibly uniform observed GC colors in DF2 and DF4 (intrinsic observed $\sigma_{(V-I)}=0.015$ mag of a combined GC sample) are thought to have formed in a collision \citep{shen2021,vandokkum2022}. In contrast, NGC~5846-UDG1 and DGSAT~1 are thought to have uniform populations due to rapid, possibly clumpy star formation at an early epoch \citep{danieli2022,janssens2022}, but similar star formation conditions may arise from a collision. 


\begin{figure}
    \centering
    \includegraphics[width=\linewidth]{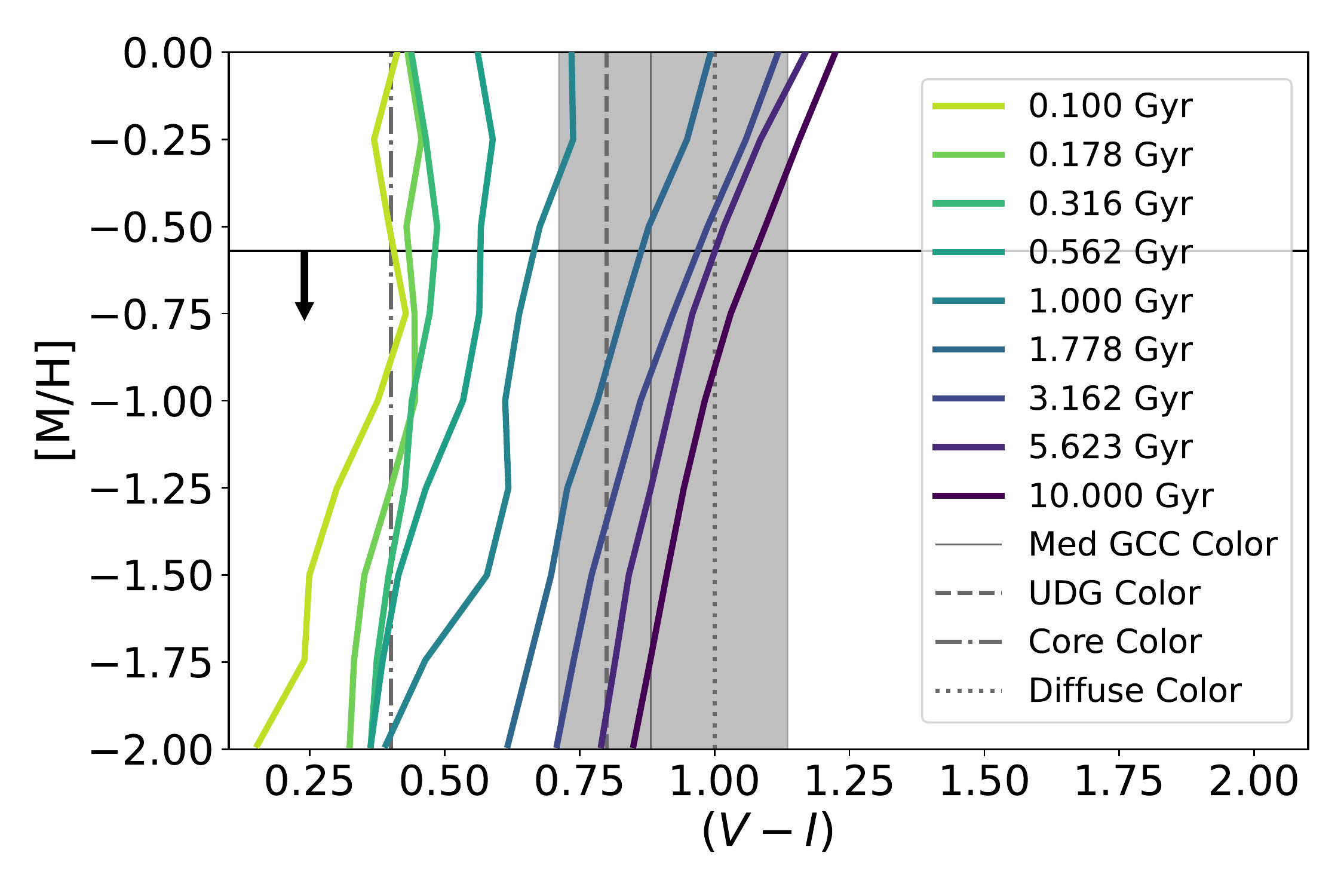}
    \caption{Metallicity [M/H] as a function of $(V-I)$ color for single-burst stellar populations of different ages with a Chabrier log-normal IMF. The gray shaded region marks the color range of our GCCs. Using the mass-metallicity relation from \citet{andrews2013} we derive an [M/H] upper limit of $-0.57$ (black horizontal line with downward arrow), which implies a minimum age of $\sim 1.5$~Gyr for our GCCs. But otherwise degeneracy between age and metallicity makes it difficult to pinpoint where in parameter space the GCCs lie. We also include vertical lines for the median GCC color and colors of the UDG itself, but do not depict the errors on these quantities for readability, although they are substantial (see \autoref{tab:udg}).}
    \label{fig:color_metallicity}
\end{figure}

To further inspect our GC candidate ages and metallicities we employ the \textsc{Parsec v.1.2S} code \citep{bressan2012}. We produce results for a single burst stellar population with a Chabrier log-normal IMF. The results of this analysis are plotted in \autoref{fig:color_metallicity}. Here we plot lines of constant age in color-metallicity space. 
If we use the mass-metallicity relation from \citet{andrews2013} and assume the ratio of oxygen to other metals is the same in the UDG as in the Sun (and using the solar Oxygen abundance from \citealt{asplund2009}) we find an upper limit on [M/H] $\approx-0.6$ for the UDG, which implies a minimum age of $\sim1.5$~Gyr for the GCCs. If the GCs formed at earlier times, which is likely, they will be lower than the Sun in metallicity. And, even if this galaxy does not follow the mass-metallicity relation, the GCs must still be older than $\sim1$~Gyr (the minimum age that falls within our GC color selection range). While spectroscopic follow-up is needed for additional constraints, this is difficult or impossible for most of the GCCs given their brightness (see \autoref{tab:gccs}).


\subsubsection{The Globular Cluster Luminosity Function}
\label{subsubsec:gclf}

\begin{figure}
    \centering
    \includegraphics[width=\linewidth]{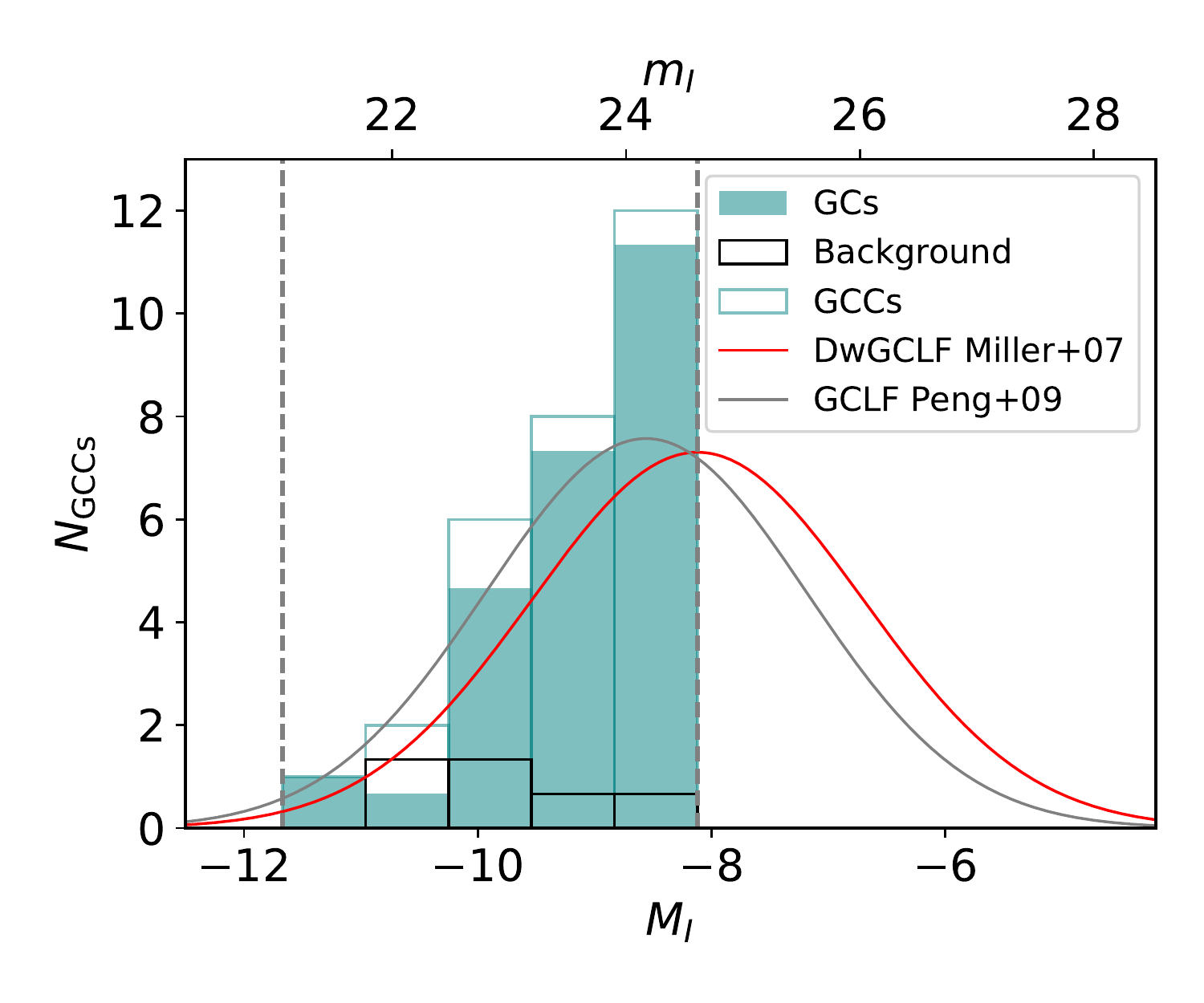}
    \caption{The observed GCLF of UGC~9050-Dw1. The 30 GCCs are plotted with outlined teal bars, and the contaminant counts normalized to sum to 4 is plotted with outlined black bars. The latter is subtracted from the former to construct our final counts. The two $I-$band magnitude limits are marked with vertical dashed lines. We compare to the GCLF determined from dwarf ellipticals by \citet{miller2007} (gray line) and the GCLF determined from M87 by \citet{peng2009} (red line). The GCLF of UGC~9050-Dw1 is well matched to the dwarf elliptical GCLF considering small number statistics.}
    \label{fig:GCLF_I}
\end{figure}

The GCLF has a near universal peak, with some variance, dependent on the general galaxy type. In \autoref{fig:GCLF_I} we plot the GCLF of UGC~9050-Dw1 with solid teal bars. Here we compare to two GCLFs: 1) the \citet{miller2007} GCLF determined from dwarf ellipticals in Virgo found to peak at $\mu_{M_{I},\rm{Vega}} = -8.12$ mag with $\sigma_{M_{I}} = 1.42$ for which the probability distribution function (PDF) is plotted in red, and 2) the \citet{peng2009} GCLF determined from M87, a giant elliptical, found to peak at $\mu_{M_{I},\rm{Vega}} = -8.56$ mag with $\sigma_{M_{I}} = 1.37$ for which the PDF is plotted in gray. 

The observed GCLF for UGC~9050-Dw1 follows the expected GCLF for dwarfs reasonably well (marginally better than the standard GCLF); the $V-$band GCLF yields similar results. Hence why we select our GCCs in magnitude space covering the bright half of the GCLF. We do the same analysis down to the $90\%$ completeness limit ($m_{F814W} = 26.8$ mag; $M_{I}=-5.55$) and again find general agreement with a standard dwarf GCLF, but with a slight overabundance at the faint end, likely due to contamination. The agreement with the GCLF also indicates that the distance we have assumed for  UGC~9050-Dw1 is approximately correct.

\subsubsection{Specific Frequency}
\label{subsubsec:specfreq}

Specific frequency offers a measurement of how rich a GC system is relative to its host galaxy luminosity. We follow the definition presented in \citet{harris1981} of $S_{N} = N_{\rm{GC}}10^{0.4(M_{V}+15)}$ in combination with our ground-based CFHT measurements and final GC count of UGC~9050-Dw1 (\autoref{tab:gc_counts}). 
This gives a specific frequency of $S_{N} = 122\pm38$. 

\autoref{fig:ngc_mv} includes diagonal lines depicting constant specific frequency of 1, 10, and 100. Other well-studied UDGs with high specific frequencies include DF17 ($28\pm5$), VCC~1287 ($80\pm29$), and NGC~5846-UDG1 ($78$ derived from \citealt{danieli2022}, $58\pm14$ in \citealt{muller2021}). Even amongst these unusual UDGs, UGC~9050-Dw1 has an extraordinary specific frequency. The high specific frequency objects in clusters ($S_{N}>100$) at present do not have evidence of recent star formation. Other star forming dwarfs, low surface brightness galaxies, and even spiral galaxies typically have much lower specific frequencies. For example \citet{forbes2020} derives a specific frequency for the Milky Way $S_{N}= 58$. 

\subsubsection{Stellar Fraction in Globular Clusters}
\label{subsubsec:gc_fraction}

Here we determine the fraction of light within the entire UDG that is contained in globular clusters. First, the total flux within the UDG globular clusters is determined by integrating the background subtracted histogram (in flux space) in $V$-band. We account for the portion of the $V$-band GCLF that we do not sample by multiplying the total flux by $\frac{10}{7}$, as we sample approximately $70\%$ of the luminosity function in $V$. Uncertainties are determined with the same approach as that presented in \citet{danieli2022}. We perturb the measured histogram values by their errors 10,000 times and then derive the confidence interval of the resulting magnitude distribution at $2\sigma$. We find that $M_{V,GCs}=-12.3\pm0.5$ mag. This implies a flux ratio of $0.21 \pm 0.1$ or that $21\% \pm 10\%$ of the stars measured in $V$-band reside in globular clusters. For comparison most galaxies fall within the range of $0.1-1\%$, with Coma UDGs showing higher fractions (in the 10\% range), and NGC~5846-UDG1 containing the highest fraction at $12.9\%\pm0.6\%$. Considering the substantial errors on UGC~9050-Dw1 it appears to be consistent with other UDGs on the high end. 


\subsubsection{The Dark Matter Halo Mass}
\label{subsubsec:dm_mass}

The abundance of GCs and the mass in GCs can both be used as an estimator of the dark matter halo mass of a system. The mass of a galaxy has been found to scale near linearly in GC abundance and total GC mass across a very large mass range \citep[e.g.,][]{el-badry2019,forbes2020}, down to $M_{\rm{total}}\sim10^{8.75}$\msun \citep{zaritsky2022}. In terms of total system mass (stellar mass + halo mass) \citet{zaritsky2022} finds 1 GC per $2.9 \pm 0.3 \times 10^{9}$\msun. With our GC abundance this relation yields $M_{\rm{total}} = 1.5\pm0.3\times10^{11}~\msun $ for UGC~9050-Dw1. Because this mass estimate is 3 orders of magnitude larger than our stellar mass estimate we make the approximation that $M_{\rm{total}} \approxeq M_{\rm{halo}}$. \citet{harris2017} finds an average GC mass of $1\times10^{5}~\msun$ for UDGs VCC~1287 and DF44, a value similar to what is expected from other dwarf measurements. Using this average mass and the relation $\frac{M_{\rm{GC}}}{M_{\rm{halo}}} = 2.9\times10^{-5}$ from \citet{harris2017} we find a halo mass of $M_{\rm{halo}} = 1.8\pm0.3\times10^{11} $\msun\ for UGC~9050-Dw1. Both approaches give consistent estimates of $M_{\rm{halo}}$.

For comparison, \citet{penarrubia2016} and \citet{erkal2019} find the mass of the LMC $M_{\rm{LMC}} = 2.5^{+0.9}_{-0.8} \times 10^{11}~\msun$, with a total of $\sim$40 GCs \citep[e.g.][and the references therein]{Baumgardt2013,bennet2022}. In general bright dwarfs are often defined to have stellar masses of up to $10^{9}~\msun$ which corresponds to halo masses in the range of a $\sim$ a few $\times10^{10}~\msun$ to $10^{11}~\msun$ \citep{behroozi2013,bullock2017} implying that UGC~9050-Dw1 may have an unusually high halo mass for its stellar mass, but within expectation for massive dwarfs. Other well-studied UDGs with presumed overly massive dark matter halos given their stellar mass include DF17 ($\sim 9\times10^{10}~\msun$), VCC~1287 ($\sim 8\times10^{10}~\msun$), and NGC~5846-UDG1 ($1.6\times10^{11}~\msun$ using the \citealt{zaritsky2022} relation and GC counts from \citealt{danieli2022}; $9\pm2\times10^{10}\msun$ in \citealt{muller2021}). 

While it appears that UGC 9050-Dw1 possesses an excessively massive dark matter halo, we want to emphasize that there are some caveats to consider in this mass estimate. Firstly, the assumed GC abundance - halo mass relation may not be applicable at the low stellar masses typical of UDGs \citep[see e.g.,][]{burkert2020,liang2023}, despite being commonly assumed in the literature. Secondly, UGC 9050-Dw1's disturbed nature raises concerns about its dynamical equilibrium, which could impact the applicability of the GC abundance - halo mass relation, even if it holds at lower masses.


\section{The Formation of UGC~9050-Dw1}
\label{sec:discussion}

In terms of formation mechanisms for the UDG UGC~9050-Dw1, there are several key observed features that must be considered:
\begin{enumerate}

    \item The optical and Apertif \hi\ morphologies of UGC~9050-Dw1 strongly suggest a past or ongoing interaction, likely linked to its current UDG appearance. This implies that we may have observed UGC~9050-Dw1 in the process of becoming a UDG.
    
    \item UGC~9050-Dw1 is an \hi-bearing UDG with bright NUV flux and corresponding blue colors indicative of recent star formation in the ``core'' region. Any interactions have not quenched the system, or must have done so recently ($< 100$~Myr).
    
    \item The GC candidate colors of UGC~9050-Dw1 are remarkably uniform and blue, suggestive of a single epoch of GC formation. This could have occurred either if the progenitor was a massive dwarf galaxy before transforming into a UDG or, more likely, during the process that transformed the progenitor into a UDG.
    
    \item The GC counts of UGC~9050-Dw1 are exceptionally high, with one of the highest observed specific frequencies and fraction of stellar light in GCs for a UDG. This implies an epoch of clumpy and high star formation density in order to form such an abundance of GCs.
    
    \item UGC~9050-Dw1 and its assumed host UGC~9050 are strikingly similar in both estimated \hi\ mass and color, with little evidence of direct interaction. 
    
\end{enumerate}

In the following subsections we comment on several of the proposed UDG formation mechanisms and how they mesh with the key criteria we note above. We favor the dwarf merger formation mechanism (\autoref{subsec:dwarf_collision}), followed by the disrupted galaxy formation mechanism (\autoref{subsec:disrupted_gal}) over other common UDG formation mechanisms.

\subsection{Less Likely Formation Origins}

\textbf{Tidal Dwarf Galaxy} These objects (TDGs), are believed to form in the debris of interacting spiral galaxies, arising from instabilities in tidal tails or ejected stellar material \citep[e.g.,][]{elmegreen1993,duc1998,kaviraj2012}. TDGs are expected to lack dark matter and globular clusters \citep[e.g.,][]{barnes1992,duc2004,bournaud2006,duc2012,ploeckinger2018} and are outliers in the mass-metallicity relation due to their formation from metal-enriched material \citep[e.g.,][]{duc2012,sweet2014,dumont2021}. Although we cannot directly determine the metallicity of UGC~9050-Dw1, \autoref{fig:color_metallicity} shows that both the GCs and the UDG itself exhibit colors consistent with metallicities similar to or slightly more metal-rich than the Sun ($\sim 0.5$ dex), indicating that UGC~9050-Dw1 is not metal rich. Additionally, TDGs are typically found close to their parent galaxies, within 15 optical half-light radii \citep{kaviraj2012}. However, UGC~9050-Dw1 is situated  43~kpc away from its parent galaxy, as determined from the $g$-band exponential disk length of UGC~9050 and the relation $R_{e}\sim1.68R_{d}$. An abundance of GCs similar to what we observe in UGC 9050-Dw1 is unprecedented for a TDG. Additionally, the absence of strong \hi\ tidal tails, typical of young TDGs (while old TDGs may have no gas), further weakens the likelihood of UGC~9050-Dw1 originating as a TDG.

\textbf{Tidally `Puffed' Dwarf} 
Tidal stripping and heating of dwarf-mass DM halos by a massive host results in an expansion of the half-light radii of the stellar component, which can result in a ``puffed up'' dwarf; i.e. a UDG \citep{carleton2019,tremmel2020}. Yet the GC abundance (and inferred halo mass, if the estimate is to be believed) of UGC~9050-Dw1 is not reflective of the average dwarf GC population, in sharp contrast to the two group UDGs with associated tidal features studied in \citet{jones2021} (containing 2 and 5 GCs). The models of \citet{carleton2021} are able to produce GC rich UDGs as a product of tidal heating, but it requires such UDGs to live in massive galaxy clusters that are tidally heated and thus ``puffed up'' during cluster in-fall, post intense star formation at high redshift. 
In addition to the GC counts, the similarities between UDG UGC~9050-Dw1 and its presumed companion UGC~9050 (in particular rather similar \hi\ masses) and relatively isolated environment indicate that the triggering mechanism for UGC~9050-Dw1 could not have been typical tidal stripping/heating.

\textbf{Failed Galaxy} \citet{peng2016} posit that some UDGs may be galaxies in which the old stellar halo was able to form, but then subsequent rapid gas removal resulted in the lack of bulge or disk formation, while a massive dark matter halo remains \citep[see also e.g.,][]{vandokkum2015,vandokkum2016}. These UDGs are called ``failed galaxies'', such that they were on the path to becoming galaxies within the mass range of the Magellanic Clouds or M33, but then this formation was interrupted. During an intense early epoch of star formation for which the gas surface density is high, high GC mass fractions arise, with the remainder of the stellar population old and metal poor \citep{ferre-mateu2018,forbes2020,trujillo-gomez2021,villaume2022}. It has also been argued that for some UDGs, like NGC~5846-UDG1 \citep{muller2021,danieli2022} and DGSAT 1 \citep{janssens2022}, a single early star formation burst may have occurred before the galaxy ``failed'', resulting in monochromatic GC populations. However, UDGs of this origin are generally thought to be devoid of cold gas as a result of 
ram pressure stripping and/or strangulation post cluster infall \citep{boselli2014,yozin2015,beasley2016,beasley2016A,safarzadeh2017,chan2018,jiang2019,tremmel2020}, or intense supernova feedback from the same star forming event responsible for the mono-GC population \citep[e.g.,][]{dicintio2017,chan2018,danieli2022} where an old and red stellar population remains. While at first glance UGC~9050-Dw1 meshes with this formation mechanism, the clear tidal features and substantial \hi\ mass in conjunction with NUV luminosity and blue ``core'' color indicative of recent star formation disfavor a failed galaxy origin.

\subsection{A Disrupted Galaxy}
\label{subsec:disrupted_gal}

In this subsection we propose a disruption event as a possible formation mechanism for UGC~9050-Dw1. Fundamentally this formation scenario is akin to a tidally ``puffed up'' dwarf, like those observed in \citet{jones2021} or \citet{zemaitis2023}. However, in this instance the progenitor of UGC~9050-Dw1 is a low mass spiral galaxy (or possibly a massive dwarf). The tail then is a result of a tidal interaction between the progenitor of UGC~9050-Dw1 and another galaxy. This means that the galaxy with which UGC~9050-Dw1 interacted may have had a similar mass (like the close, equal mass passage explored in \citealt{toomre1972}) or been more massive. And, in this instance, the galaxies did \textit{not} collide.  

There are pieces of evidence that support the proposition of the UGC~9050-Dw1 progenitor falling into the category of a ``normal'' galaxy, aside from the distribution of stars and present day morphology (such as the clumpy star formation). First, from the \hi\ perspective UGC~9050-Dw1 may not be as unusual as it seems. The \hi\ mass to stellar mass ratio is $\approx 8.7$, which is comparable to gas bearing dwarfs in the field (typical ratios are 1 to 10, see e.g., fig.2 of \citealt{huang2012}). In contrast, \hi\ bearing UDGs have been found to be more \hi\ rich, with somewhat higher ratios (\citealt{leisman2017}; see fig. 4). It is important to note that other UDGs in the field with similar \hi\ masses to our UDG have almost no GCs \citep{jones2023}, indicative of typical dwarf mass progenitors. Galaxy disruption events may trigger GC formation which could explain the GC abundance and relatively monochromatic colors of the GCs resulting from the single burst, but it is unclear exactly how efficient this mechanism is at forming a high volume of GCs in these relatively lower mass systems.

However, there are a few important caveats to consider for a massive dwarf/lower mass spiral progenitor. The halo mass estimate of UGC~9050-Dw1 is comparable to the LMC. In order to disrupt the older stellar population of the LMC ($M_{*}=2.7\times10^{9}\msun$; \citealt{vandermarel2002}) enough for it to become a UDG (and assuming a similar initial stellar mass for UGC~9050-Dw1) almost $90\%$ of the stars have been blown out and become nearly undetectable while the \hi\ remains roughly intact, which is not a readily explainable interaction. Work by \citet{benavides2021} finds that backsplash UDGs are entirely stripped of their gas via ram pressure stripping, which further necessitates the need for an approximate equal mass passage. 

Second, it is not completely clear what UGC~9050-Dw1 would have interacted with to have disrupted. While UGC~9050 has a comparable \hi\ mass, there is only a hint of possible disturbance in \hi\ as seen in \autoref{fig:host_radio_img} and none in the optical. Typically we would expect to see a similar magnitude of disruption in UGC~9050, although it is possible that the dynamics may have been such that UGC~9050 was minimally impacted. It is plausible that UGC~9050-Dw1 interacted with the nearby ($\sim 300$ kpc) NGC~5481 group instead of UGC~9050. Such an interaction would have happened a $\sim$few Gyrs ago to explain the current location of UGC~9050-Dw1, but this passage would again be akin to the mass differential between the LMC and Milky Way. It is possible UGC~9050-Dw1 may instead have interacted with a galaxy that is no longer detectable due to strong disruption, somewhat akin to the dwarf collision discussed below.

Last, it is not clear how a monochromatic GC population would arise from a disruption event. Disruption events can lead to star formation bursts, but typically then we would expect a clear older GC component remaining from the progenitor. For example, the LMC has young and blue clusters like those observed in UGC~9050-Dw1 but also contains many old and red GCs. If the GCs formed prior to disruption, their uniform color is still difficult to explain, as dwarf and spiral galaxies tend to have a larger spread in color (see \autoref{fig:color_dist} and discussion), but may be plausible.  



\subsection{Dwarf Collision} 
\label{subsec:dwarf_collision}

The second possible mechanism for the formation of UGC~9050-Dw1 is that of a dwarf collision/major merger, which we argue is the most plausible formation mechanism. In contrast to the galaxy disruption formation scenario where two objects had a close encounter, here two galaxies collided, and UGC~9050-Dw1 is the resulting amalgamation of the two. 
The general morphology of the tail seen in UGC~9050-Dw1 is not unlike that observed in dwarf mergers \citep{paudel2018,micic2023}.  
In UGC~9050-Dw1 the tail is large enough that we suspect such a merger would have been approximately equal mass with a relatively small impact parameter - i.e., a merger not cataclysmic enough such that much of the gas has remained intact and a tail is produced, with the remnant consisting of both parent objects once the perturber falls in.

\citet{wright2021} studied the formation of UDGs in low density environments in hydrodynamical simulations (ROMULUS25), finding that a majority of UDGs were produced as a result of major mergers at early times in the Universe. These mergers increase the effective radius and angular momentum of the progenitor, ultimately decreasing the central surface brightness and re-distributing star formation to the outskirts. Such an origin for UGC~9050-Dw1 implies that this may be a more recent merger ($\lesssim 10$~Gyr, depending on the age of the GCs), before the formation of the steep color gradient between a red center and blue outskirts that has been seen in simulations. While \citet{wright2021} found that massive UDGs tend to have their star formation de-centralized, this trend was not as clear for lower mass UDGs, where UGC~9050-Dw1 falls into the low mass regime of the simulated ROMOULUS25 UDGs. It is also possible that instead the UDG is not face-on, and actually the star forming clump is in the outskirts and only projected to be at the center. The Apertif DR1 data in \autoref{fig:radio_img} suggests the clump may be offset from the gas, but the apparent gas distribution may be significantly impacted by noise and the beam side-lobes making it difficult to constrain viewing angle.

For a dwarf major merger scenario to have occurred, we must explain how a high abundance of monochromatic GCs could form. Upon investigating a collision as a UDG origin for DF2 and DF4, which \citet{vandokkum2022} invoked to explain the highly monochromatic GC populations, \citet{ogiya2022} find that as many as 30-60 GCs can be produced in a single burst. Studies of other dwarf mergers also find subsequent high star formation rates \citep{paudel2018,kado-fong2020,zhang2020,egorova2021}. Therefore it is plausible that here in the immediate aftermath of the merger there was an episode of high density star formation that spurred the GC formation. Rapid, extensive GC formation requires this to be a ``wet'' merger, in which case the \hi\ we see is the final remnant of gas from the pair. 

The dwarf merger formation theory is further supported by the high luminosity fraction of GCs we find for UGC~9050-Dw1, which is comparable to UDGs in the Coma cluster and NGC~5846-UDG1. \citet{danieli2022} argues that NGC~5846-UDG1 results from extreme conditions causing clumpy star formation yielding a high fraction of stars to form as GCs. Collisions have been found to cultivate unusual/clumpy star formation conditions, even in the dwarf regime \citep[e.g.,][]{kado-fong2020,lahen2020,zhang2020,kimbro2021,gao2022}, making it very plausible that the high GC abundance observed in UGC~9050-Dw1 is a result of a galaxy merger. This short lived, intense episode of star formation also neatly explains the monochromatic GC color we observe in which the GCs all formed at approximately the same time. We propose that, while these events may be less common than other UDG formation mechanisms, dwarf mergers may serve as an avenue to form GC rich UDGs with relatively monochromatic GC populations. 

Naively, the lack of an \hi\ tail tracing the stellar tail may oppose a dwarf collision origin, as the standard theory is that \hi\ tails are much longer lived (order of Gyr) than stellar tails (order of 100 Myrs; e.g., see review by \citealt{duc2013}). However, at lower surface brightness this is difficult to observe. A merger would pull both gas and stars into the same tail, but low column-density \hi \ would eventually be photoionised away from the background UV radiation field. Our \hi\ data is also rather shallow, so any \hi\ in the tail may merely be below our detection limit.

If this system formed via a merger, we can estimate whether we expect the UDG to be in dynamical equilibrium. We have loose constraints on the age of the GCs, spanning from 1.5 to 10 Gyrs, which poses challenges in precisely pinpointing when the merger occurred. Typically, the signatures of a merger tend to fade after several Gyrs (2-5 Gyrs see e.g., \citealt{johnston2001,pearson2018}). A case study by \citet{tarumi2021} suggests that effective relaxation after a merger may take around 2 Gyr to initiate. Moreover, simulations of TDGs by \citet{bournaud2006} reveal similar timescales, with self-gravitating TDGs lasting for up to a few Gyrs. Based on this information, we can infer that the relaxation time for this UDG likely falls between 1-5 Gyrs, with a dynamical time of a few hundred Myr. Given that the tail is still prominently visible, we speculate that the merger likely transpired within the past $\sim$3 Gyr or even more recently. Consequently, it is improbable that the UDG is presently in dynamic equilibrium.


\bigskip
In either the disrupted galaxy or dwarf collision formation scenario, the red colors of the ``diffuse'' component of UGC~9050-Dw1 could be the remnant of an old halo component from the progenitor(s).

\section{Conclusion}
\label{sec:conclusion}

In this paper we investigate the disturbed, \hi\ bearing UDG UGC~9050-Dw1 and its GC system using HST/ACS F555W and F814W filters and the VLA D-array. UGC~9050-Dw1 was thought to be a companion to the low surface brightness spiral UGC~9050, lying $\sim70$~kpc to the West of UGC~9050-Dw1.
UGC~9050-Dw1 contains a central, elongated, UV bearing, blue core with an extended tail feature and a redder diffuse component. We find a total GC abundance of $52\pm12$ with a specific frequency of $S_{N} = 122\pm38$ and GC luminosity fraction of $21\%\pm10\%$, marking an extreme in UDG parameter space. The GC colors are surprisingly uniform and on the blue end of the GC color distribution, indicative of a possible epoch of intense star formation in which a majority of the GCs formed. We estimate the GC population to be a minimum of $\sim$1.5~Gyr old. The GC population of UGC~9050-Dw1 has a centrally peaked radial distribution. Likewise the GCLF of UGC~9050-Dw1 is consistent with that expected for dwarf galaxies at the presumed distance of UGC~9050-Dw1. The VLA data indicate that the \hi\ mass of the system ($10^{8.44}\msun$) is similar to the nearby low surface brightness galaxy UGC~9050 ($10^{8.94}\msun$). Because we are at the resolution limit of the VLA D-array configuration we cannot reliably constrain the dynamical mass of the system (which may indeed not even be in dyamical equilibrium). Instead we estimate the dark matter halo mass from GC counts, and infer a massive dark matter halo ($M_{\rm{halo}}=1.5\pm0.3\times10^{11}\msun$) for its stellar mass ($M_{*}\sim10^{8}\msun$), although this relation may not apply well to UDGs and we caution that this halo mass estimate is likely incorrect. 

UGC~9050-Dw1 is truly an enigmatic object, with a number of unique properties that in combination appear peculiar for a UDG and at first glance thwart easy explanation. UGC~9050-Dw1 does not neatly fit into any of the standard proposed UDG formation scenarios while simultaneously easily classifying as a UDG within the standard definition. Most notably, in contrast to the two group UDGs with disturbed features studied in \citet{jones2021}, UGC~9050-Dw1 does \textit{not} mesh as well with the tidally heated dwarf progenitor channel of UDG formation, given its GC abundance and gas content. Instead we find UGC~9050-Dw1 to be more consistent with a disrupted galaxy or dwarf major merger as its progenitor (we favor the latter). Clumpy, high star formation density induced by a dwarf major merger more easily explains the observed characteristics of UGC~9050-Dw1, especially if the GC abundance - halo mass relation does not hold for this system and the dark matter halo is \textit{not} overly massive. 

UGC~9050-Dw1 is an important object, potentially providing us insights into how UDGs form. Specifically, we propose dwarf major mergers may be another avenue for UDG formation, especially those with high GC abundance and monochromatic GC populations, without having to invoke complex or unknown exotic processes. That said, we expect that this mechanism alone cannot explain all GC-abundant UDGs. UGC~9050-Dw1 may be one of the first observational analogues to the dwarf major merger UDGs produced in the ROMULUS25 simulation \citep{wright2021}. 
Constraints on the dynamics and metallicity of the system may provide further insight into just how this unusual galaxy came to be.

\begin{acknowledgments}

The authors would like to thank both Anil Seth and Anna Wright for useful discussions on the nature of this system. We also thank Steven Janssens for providing details on the properties of DGSAT 1.

This work is based on observations made with the NASA/ESA Hubble Space Telescope, obtained at the Space Telescope Science Institute, which is operated by the Association of Universities for Research in Astronomy, Inc., under NASA contract NAS5-26555.  These observations are associated with program \# 16890

The work used data observed with the Karl G. Jansky Very Large Array. The National Radio Astronomy Observatory is a facility of the National Science Foundation operated under cooperative agreement by Associated Universities, Inc.

This work is based on observations obtained with MegaPrime/MegaCam, a joint project of CFHT and CEA/IRFU, at the Canada-France-Hawaii Telescope (CFHT) which is operated by the National Research Council (NRC) of Canada, the Institut National des Science de l'Univers of the Centre National de la Recherche Scientifique (CNRS) of France, and the University of Hawaii. This work is based in part on data products produced at Terapix available at the Canadian Astronomy Data Centre as part of the Canada-France-Hawaii Telescope Legacy Survey, a collaborative project of NRC and CNRS.

AK acknowledges financial support from the grant CEX2021-001131-S funded by MCIN/AEI/ 10.13039/501100011033 and from the grant POSTDOC\_21\_00845 funded by the Economic Transformation, Industry, Knowledge and Universities Council of the Regional Government of Andalusia. KS acknowledges support for the Natural Sciences and Engineering Research Council of Canada (NSERC). Research by DC is supported by NSF grant AST-1814208.

\end{acknowledgments}


\vspace{5mm}
\facilities{HST(ACS), VLA, GALEX, CFHT}


\software{Astropy \citep{astropy2013,astropy2018,astropy2022}, 
          CASA \citep{mcmullin2007},
          CGAT-core \citep{CGAT-core},
          Dolphot \citep{dolphin2000,dolphin2016},
          Galfit \citep{peng2010},
          Photutils \citep{bradley2022},
          Reproject \citep{robitaille2020},
          Source Extractor \citep{bertin1996},    
          }



\appendix
\restartappendixnumbering
\section{Supplementary Tables and Figures}
\label{sec:appendix}

\begin{figure}
    \centering
    \includegraphics[width=\linewidth]{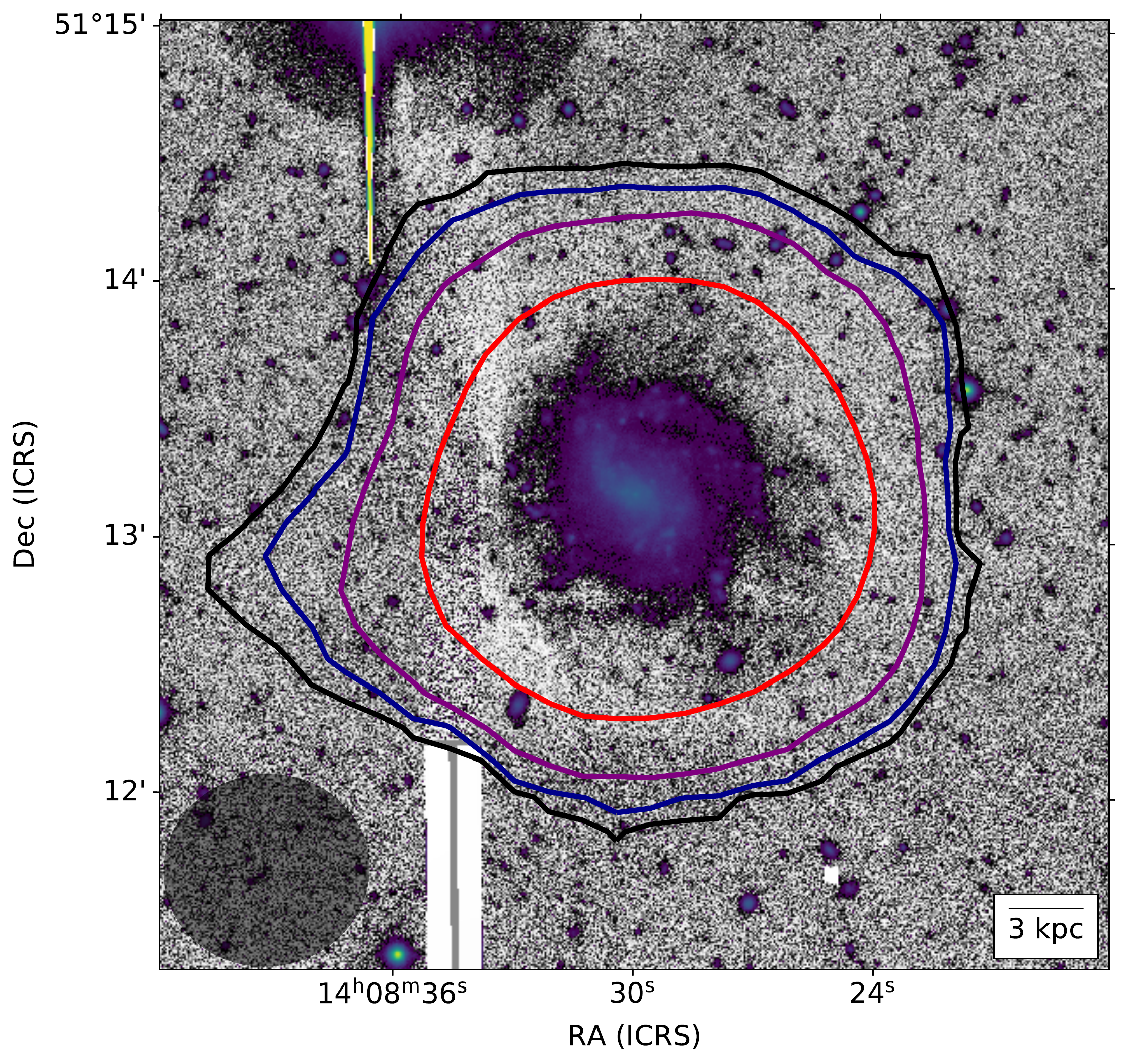}
\caption{Same as the left panel of \autoref{fig:radio_img}, but for UGC~9050, constructed from the W3$-$2$-$3 field. Image artifacts are the result of the galaxy lying at the edge of the survey frame. The minimum contour is the same as that in \autoref{fig:radio_img} ($2.7\times10^{19} \; \mathrm{cm^{-2}}$ over 20~\kms), but each subsequent contour is triple the previous one instead of double for ease of readability. There is a slight HI extension to the east in the direction of the presumed companion UGC~9050-Dw1.}
\label{fig:host_radio_img}
\end{figure}

Here we provide additional details on the presumed host galaxy, UGC~9050, and on the individual GCCs of UGC~9050-Dw1. For UGC~9050, \autoref{tab:host} summarizes the measured properties and \autoref{fig:host_radio_img} presents the VLA D-array results.
Lastly, \autoref{tab:gccs} presents detailed information on each of the 30 GCCs identified in this study.

\begin{table}[]
\centering
\caption{UGC~9050 Properties}
    \begin{tabular}{r|l}
    \hline\hline
         RA (J2000) & 14:08:30.056 \\
         Dec (J2000) & +51:13:10.88 \\
         $D$ (Mpc) & $35.18\pm2.48$ \\
         $M_{V}$ & -$16.7\pm0.1$ \\
         $(V-I)$ & $0.8\pm0.1$ \\
         $m_{g}$ & $16.2\pm0.1$ \\
         $(g-r)$ & $0.4\pm0.1$ \\
         $m_{NUV}$ & $17.3\pm0.1$ \\
         $\log{M_{*}/M_{\odot}}$ & $8.57\pm0.1$ \\
         $\log{M_{\hi\ }/M_{\odot}}$ & $8.94\pm0.04$ \\
         SFR (\msun\ yr$^{-1}$) & $0.056\pm0.005$ \\
    \hline
    \end{tabular}\\ [4pt]
    \label{tab:host}
\end{table}


\begin{table*}[]
\centering{}
\caption{Globular cluster candidates in UGC~9050-Dw1}
    \begin{tabular}{cccccccccccc}
    \hline\hline
    No. & R.A. & Dec &  F555W  &  F814W & $m_{V}$ & $m_{I}$ &  $M_{I}$ & $(V-I)$ & $c_{4-8}$ & $\rm{D}_{\rm{cen}}$ \\
     & deg & deg & mag & mag & mag & mag & mag & mag & & kpc \\
    \hline

    1  & 212.305373 &  51.242282 &  $22.18 \pm 0.01$ &  $20.68 \pm 0.01$ &  22.10 &  20.68 & -10.63 &      1.43 &           0.41 &     10.96 \\
    2  &  212.284904 &  51.230894 &  $22.64 \pm 0.01$ &  $21.79 \pm 0.01$ &  22.59 &  21.70 & -10.14 &      0.90 &           0.52 &      8.40 \\
    3  &  212.301603 &  51.243946 &  $22.93 \pm 0.01$ &  $21.98 \pm 0.01$ &  22.88 &  21.97 &  -9.85 &      0.91 &           0.59 &     12.01 \\
    4  &  212.301313 &  51.218302 &  $23.12 \pm 0.01$ &  $21.82 \pm 0.01$ &  23.06 &  21.82 &  -9.68 &      1.24 &           0.39 &      3.93 \\
    5  &  212.319245 &  51.223368 &  $23.88 \pm 0.02$ &  $22.96 \pm 0.01$ &  23.84 &  22.958 &  -8.90 &      0.88 &           0.56 &      5.84 \\
    6  &  212.298561 &  51.220217 &  $23.81 \pm 0.02$ &  $22.85 \pm 0.01$ &  23.76 &  22.849 &  -8.98 &      0.91 &           0.58 &      3.37 \\
    7  &  212.303981 &  51.237101 &  $23.85 \pm 0.01$ &  $22.96 \pm 0.01$ &  23.81 &  22.96 &  -8.93 &      0.85 &           0.54 &      7.77 \\
    8  &  212.286567 &  51.213581 &  $23.84 \pm 0.01$ &  $22.91 \pm 0.01$ &  23.80 &  22.90 &  -8.94 &      0.89 &           0.59 &      9.50 \\
    9  &  212.299176 &  51.218954 &  $23.93 \pm 0.02$ &  $22.97 \pm 0.01$ &  23.88 &  22.96 &  -8.85 &      0.92 &           0.55 &      3.88 \\
    10 &  212.310203 &  51.215676 &  $23.98 \pm 0.02$ &  $23.09 \pm 0.01$ &  23.93 &  23.09 &  -8.80 &      0.84 &           0.62 &      5.86 \\
    11 &  212.312055 &  51.221985 &  $24.29 \pm 0.02$ &  $23.27 \pm 0.01$ &  24.24 &  23.27 &  -8.50 &      0.97 &           0.62 &      3.39 \\
    12 &  212.320992 &  51.221122 &  $24.34 \pm 0.02$ &  $23.36 \pm 0.01$ &  24.30 &  23.36 &  -8.44 &      0.93 &           0.55 &      6.78 \\
    13 &  212.315175 &  51.222245 &  $24.32 \pm 0.02$ &  $23.43 \pm 0.01$ &  24.27 &  23.42 &  -8.46 &      0.85 &           0.62 &      4.44 \\
    14 &  212.326368 &  51.218505 &  $24.34 \pm 0.02$ &  $23.75 \pm 0.02$ &  24.31 &  23.74 &  -8.43 &      0.56 &           0.72 &      9.28 \\
    15 &  212.302150 &  51.236079 &  $24.62 \pm 0.02$ &  $23.69 \pm 0.01$ &  24.58 &  23.69 &  -8.16 &      0.89 &           0.65 &      7.19 \\
    16 &  212.306140 &  51.230121 &  $24.69 \pm 0.02$ &  $23.79 \pm 0.02$ &  24.64 &  23.79 &  -8.09 &      0.85 &           0.67 &      3.57 \\
    17 &  212.326890 &  51.211341 &  $25.19 \pm 0.03$ &  $23.82 \pm 0.02$ &  25.12 &  23.81 &  -7.61 &      1.31 &           0.73 &     11.88 \\
    18 &  212.291674 &  51.213666 &  $24.79 \pm 0.03$ &  $23.86 \pm 0.02$ &  24.74 &  23.86 &  -7.99 &      0.88 &           0.54 &      8.18 \\
    19 &  212.292061 &  51.206527 &  $24.76 \pm 0.03$ &  $23.98 \pm 0.02$ &  24.72 &  23.976 &  -8.01 &      0.74 &           0.68 &     11.95 \\
    20 &  212.299931 &  51.225688 &  $24.73 \pm 0.03$ &  $24.00 \pm 0.02$ &  24.69 &  24.00 &  -8.04 &      0.69 &           0.61 &      1.80 \\
    21 &  212.293856 &  51.224565 &  $24.66 \pm 0.03$ &  $23.92 \pm 0.02$ &  24.62 &  23.92 &  -8.11 &      0.70 &           0.55 &      3.97 \\
    22 &  212.306118 &  51.230356 &  $24.83 \pm 0.03$ &  $24.02 \pm 0.02$ &  24.79 &  24.02 &  -7.95 &      0.77 &           0.63 &      3.71 \\
    23 &  212.310482 &  51.229573 &  $24.91 \pm 0.03$ &  $23.99 \pm 0.02$ &  24.86 &  23.98 &  -7.87 &      0.87 &           0.52 &      3.98 \\
    24 &  212.301336 &  51.219762 &  $25.36 \pm 0.05$ &  $23.97 \pm 0.02$ &  25.29 &  23.97 &  -7.44 &      1.33 &           0.42 &      3.07 \\
    25 &  212.299695 &  51.226278 &  $25.08 \pm 0.03$ &  $24.14 \pm 0.02$ &  25.04 &  24.14 &  -7.70 &      0.90 &           0.54 &      2.06 \\
    26 &  212.303734 &  51.223784 &  $24.93 \pm 0.03$ &  $24.23 \pm 0.02$ &  24.89 &  24.23 &  -7.84 &      0.67 &           0.40 &      0.44 \\
    27 &  212.324322 &  51.213237 &  $25.09 \pm 0.03$ &  $24.21 \pm 0.02$ &  25.05 &  24.20 &  -7.69 &      0.84 &           0.65 &     10.37 \\
    28 &  212.276346 &  51.225082 &  $25.13 \pm 0.03$ &  $24.25 \pm 0.02$ &  25.09 &  24.25 &  -7.64 &      0.84 &           0.58 &     10.71 \\
    29 &  212.303447 &  51.223879 &  $25.94 \pm 0.07$ &  $24.42 \pm 0.02$ &  25.87 &  24.417 &  -6.87 &      1.45 &           0.43 &      0.44 \\
    30 &  212.291680 &  51.242439 &  $25.44 \pm 0.04$ &  $24.48 \pm 0.03$ &  25.39 &  24.48 &  -7.34 &      0.91 &           0.72 &     12.05 \\
    \hline
    \end{tabular}\\[4pt]
    Columns: 1) Number. 2) Right ascension in decimal degrees. 3) Declination in decimal degrees. 4) F555W apparent magnitude and errors determined from $\textsc{Dolphot}$. 5) F814W apparent magnitude and errors determined from $\textsc{Dolphot}$. 6) $V$-band extinction corrected apparent magnitude determined from conversions presented in \citet{sirianni2005}. 7) $I$-band extinction corrected apparent magnitude. 8) $I$-band extinction corrected absolute magnitude, determined using the distance of UGC~9050-Dw1. 9) $(V-I)$ color. 10) Concentration index determined from the GCC magnitudes in 4 and 8 pixel diameter apertures. 11) Projected distance from the UGC~9050-Dw1 center in kpc.
    \label{tab:gccs}
\end{table*}



\bibliography{refs}{}
\bibliographystyle{aasjournal}



\end{document}